\documentclass{article}

\PassOptionsToPackage{numbers, compress}{natbib}

\usepackage[final]{neurips_2023}

\usepackage[utf8]{inputenc} 
\usepackage[T1]{fontenc}    
\usepackage{hyperref}       
\usepackage{url}            
\usepackage{booktabs}       
\usepackage{amsfonts}       
\usepackage{nicefrac}       
\usepackage{microtype}      
\usepackage[table,xcdraw]{xcolor}         
\usepackage{enumitem}  

\usepackage[ruled,linesnumbered]{algorithm2e}

\usepackage{natbib}

\usepackage{graphicx}
\graphicspath{{figures/}}

\usepackage{stmaryrd}
\usepackage{comment}

\usepackage{amsmath}
\usepackage{amssymb}
\usepackage{mathtools}
\usepackage{amsthm}

\theoremstyle{plain}
\newtheorem{theorem}{Theorem}

\newtheorem{lemma}{Lemma}

\theoremstyle{definition}
\newtheorem{definition}{Definition}

\theoremstyle{remark}

\usepackage{thm-restate}
\usepackage{todonotes}

\newcommand{\PN}[1]{}
\newcommand{\iPN}[1]{}

\newcommand{\Comment}[1]{\textcolor{blue}{\textbf{\texttt{// #1}}}}

\usepackage{tocloft}

\makeatletter
\let\c@lofdepth\relax
\let\c@lotdepth\relax
\makeatother

\usepackage{subfig}

\usepackage{setspace}
\newcommand{\listappendicesname}{Appendices}
\newlistof{appendices}{apc}{\listappendicesname}
\newcommand{\appendices}[1]{\addcontentsline{apc}{appendices}{{\color{black}\textbf{\protect\numberline{\thesection}#1}}}}
\newcommand{\newappendix}[1]{\section{#1}\appendices{#1}}

\newcommand{\newsubappendix}[1]{%
  \subsection{#1}%
  \addcontentsline{apc}{appendices}{{\color{black}\hspace{1.5em}\protect\numberline{\thesection.\arabic{subsection}}\hspace{0.5em} #1}}%
}

\title{False Discovery Proportion control for aggregated Knockoffs}

\author{%
  Alexandre Blain\\
  INRIA\\
  Université Paris-Saclay\\
  \texttt{alexandre.blain@inria.fr} \\
  \And
  Bertrand Thirion \\
  INRIA\\
  CEA\\
  \texttt{bertrand.thirion@inria.fr}\\
  \AND
  Olivier Grisel \\
  INRIA\\
  \texttt{olivier.grisel@inria.fr}\\
  \And
  Pierre Neuvial \\
  Institut de Math\'ematiques de Toulouse\\
  Université de Toulouse\\
  \texttt{pierre.neuvial@math.univ-toulouse.fr}\\
}

\begin{document}

\maketitle

\begin{abstract}
Controlled variable selection is an important analytical step in various scientific fields, such as brain imaging or genomics. In these high-dimensional data settings, considering too many variables leads to poor models and high costs, hence the need for statistical guarantees on false positives. Knockoffs are a popular statistical tool for conditional variable selection in high dimension. However, they control for the expected proportion of false discoveries (FDR) and not their actual proportion (FDP). We present a new method, KOPI, that controls the proportion of false discoveries for Knockoff-based inference. The proposed method also relies on a new type of aggregation to address the undesirable randomness associated with classical Knockoff inference. We demonstrate FDP control and substantial power gains over existing Knockoff-based methods in various simulation settings and achieve good sensitivity/specificity tradeoffs on brain imaging and genomic data.
\end{abstract}

\section{Introduction}

Statistically controlled variable selection arises in many different application fields, when the aim is to identify variables that are important for predicting an outcome of interest. For instance, in the context of brain imaging, practitioners are interested in finding which brain areas are relevant for predicting behavior or brain diseases. Such problems also appear in genomics, where practitioners wish to select genes associated with disease outcomes.

More precisely, we consider here \textit{conditional} variable selection, meaning that we wish to select variables that are relevant to predict an outcome \textit{given} the other variables. This type of inference is substantially more challenging than marginal inference, especially in high-dimensional settings, where the number of variables exceeds the number of samples. 
This is typically the case for brain mapping studies that comprise at most a few hundred subjects (hence, samples), while modern functional Magnetic Resonance Imaging (MRI) scans consist of  more than 100k \textit{voxels}. In the context of conditional inference, those are typically reduced to a few hundreds of brain regions, still possibly more than the number of samples.

Importantly, statistical guarantees are needed to ensure that the inference is reliable - i.e. that the proportion of false discoveries made by the variable selection procedure is controlled.

In the Knockoffs framework \citep{barber2015controlling, candes2018panning}, this problem is tackled by building noisy copies of the original variables. These copies are then compared to their original counterpart to perform variable selection. 
The intuition underlying Knockoffs is that irrelevant variables do not get a larger weight than their Knockoff, while relevant variables do. 
Crucially, the Model-X Knockoffs procedure \citep{candes2018panning} controls the False Discovery Rate \citep{benjamini1995controlling} which is the expected proportion of false discoveries.

A major caveat with this procedure is the random nature of the Knockoffs generation process: for two runs of the Knockoffs procedure on the same data, different Knockoffs will be built and subsequently different variables may be selected. This undesirable behavior hinders reproducibility.
A second caveat is that False Discovery Rate (FDR) control does not imply False Discovery Proportion (FDP) control \citep{korn2004controlling}. This leads to potentially unreliable inference: single runs of the method can produce a much higher proportion of False Discoveries than the chosen FDR level.

In this work, we propose a novel Knockoff-based inference procedure that addresses both concerns while offering power gains over existing methods, for no significant computation cost. 
The paper is organized as follows. After a refresher on Knockoff inference and aggregation, we consider the $\pi$ statistic introduced in \cite{nguyen2020aggregation} to rank variables by relevance. 
Using the symmetry of knockoffs under the null hypothesis, we construct explicit upper bounds on the Joint Error Rate (JER; \citealp{blanchard2020post}) of these statistics, leading to FDP control.
We then use the calibration principle of  \cite{blanchard2020post} to obtain sharper bounds.
Finally, we obtain a robust version of this method using harmonic mean aggregation of the $\pi$ statistics across multiple Knockoffs draws.
We demonstrate empirical power gains in various simulation settings and show the practical benefits of the proposed method for conditionally important region identification on fMRI and genomic datasets. 

\section{Related work}
There has been much effort in the statistical community to achieve derandomized Knockoff-based inference. \cite{ren2021derandomizing} introduced the idea of running Model-X Knockoffs \citep{candes2018panning} multiple times and computing for each the proportion of runs for which it was selected. \cite{gimenez2019improving} explore the idea of sampling multiple Knockoffs simultaneously. This induces a massive computational cost, which is prohibitive compared to methods that can support parallel computing.
\cite{nguyen2020aggregation} introduced an aggregation method that relies on viewing Model-X Knockoffs as a Benjamini-Hochberg (BH) procedure \citep{benjamini1995controlling} on so-called \emph{intermediate $p$-values}. Such $p$-values can be computed on different Knockoff runs and aggregated using quantile aggregation \citep{meinshausen2009p} -- then, BH is performed on the aggregated $p$-values to select variables. This approach relies on the heavy assumption that Knockoff statistics are i.i.d. under the null. Additionally, it is penalized by the conservativeness of the quantile aggregation scheme. Alternative aggregation schemes such as the harmonic mean \cite{wilson2019harmonic} can be used but do not yield valid $p$-values.

\cite{ren2022derandomized} introduced an alternative aggregation procedure where Model-X Knockoffs are viewed as an e-BH procedure \citep{wang2022false} on well-defined e-values \citep{vovk2021values}. Since the mean of two e-values remains an e-value, aggregation is done by averaging e-values across different Knockoffs draws. Then, e-BH is performed on the aggregated e-values to select variables. FDR control on aggregated Knockoffs is achieved without any additional assumption compared to Model-X Knockoffs. 
However, this method requires the difficult setting of a hyperparameter related to the chosen risk level, which highly impacts power in practice. 
Other recent developments in Knockoffs include the conditional calibration framework of \citet{luo2022improving} which aims at improving the power of Knockoffs-based methods.

There have been a few attempts at controlling other type 1 errors than the FDR using Knockoffs. \cite{janson2016familywise} achieves k-FWER control and proposes that FDP control can be obtained by using a procedure that leverages joint k-FWER control. Recently, \cite{li2022simultaneous} introduced such a procedure to reach FDP control based on the k-FWER control introduced in \cite{janson2016familywise}.
In summary, the KOPI approach is the first one that aims at controlling the FDP of knockoffs-based inference for any aggregation scheme, leading to both accurate FDP control and increased sensitivity. 

\section{Refresher on Knockoffs}
\label{sec:knockoffs}
\textbf{Notation.} We denote vectors by bold lowercase letters. A
vector $\mathbf{x} = \{x_1, \ldots, x_p\}$ from which we removed the $j^{th}$ coordinate is
denoted by $\mathbf{x}_{-j}$, i.e. $ \mathbf{x} 
\setminus \{x_j\}$. Independence between two random vectors
$\mathbf{x}$ and $\mathbf{y}$ is denoted by $\mathbf{x} \perp
\mathbf{y}$. 
For two vectors $\mathbf{x}$ and $
\mathbf{\tilde{x}}$ and a subset $S$ of indices,  $(\mathbf{x},
\mathbf{\tilde{x}})_{swap(S)}$ denotes the vector obtained from
$(\mathbf{x}, \mathbf{\tilde{x}})$ by swapping the entries $x_j$ and
$\tilde{x}_j$ for each $j \in S$. 
Matrices are denoted by bold
uppercase letters, the only exception being the vector of Knockoff
statistics that we denote by $\mathbf{W}$ as in
\cite{barber2015controlling, candes2018panning}. 
For any set $S$, $|S|$ denotes the cardinality of $S$. 
For a vector $\mathbf{z}=(z_j)_{1 \leq j \leq p}$ and $S \subset \llbracket p \rrbracket$, we denote by $z_{(j:S)}$ (or $z_{(j)}$ when there is no ambiguity) the $j^{th}$ smallest value in the sub-vector $(\mathbf{z}_s)_{s \in S}$. 
For an integer $k$, $\llbracket k \rrbracket$ denotes the set $\{1, \dots, k\}$. Equality in distribution is denoted by $\stackrel{d}{=}$.

\textbf{Problem setup.} The input data are denoted by $ \mathbf{X} \in \mathbb{R}^{n \times p}$, where $n$ is the number of samples and $p$ the number of variables. The outcome of interest is denoted by $\mathbf{y} \in \mathbb{R}^{n}$. 
The goal is to select variables that are relevant with regards to the outcome \emph{conditionally on all others}. Formally, we test simultaneously for all $j \in \llbracket p\rrbracket$:
\begin{align}
H_{0, j}: y \perp x_j | \mathbf{x}_{-j} 
\quad \text{versus} \quad 
H_{1, j}: y \not\perp x_j | \mathbf{x}_{-j}.\nonumber
\end{align}
The output of a variable selection method is a rejection set $\hat{S}
\subset \llbracket p\rrbracket$ that estimates the true unknown support
$\mathcal{H}_{1} = \{j : y \not\perp x_j | \mathbf{x}_{-j} \}$. Its
complement is the set of true
null hypotheses $\mathcal{H}_0 = \{j : y \perp x_j | \mathbf{x}_{-j}
\}$.
Its cardinality $|\mathcal{H}_0|$ is denoted by $p_0$.
To ensure reliable inference, our aim is to provide a statistical guarantee on the proportion of False Discoveries in $\hat{S}$. The False Discovery Proportion (FDP) and the False Discovery Rate (FDR) \cite{benjamini1995controlling} are defined as:
\begin{align}
\mathrm{FDP}(\hat{S})=\frac{|\hat{S} \cap \mathcal{H}_{0}|}{|\hat{S}| \vee 1}, \quad \mathrm{FDR}(\hat{S})=\mathbb{E}[\mathrm{FDP}(\hat{S})]=\mathbb{E}\left[\frac{|\hat{S} \cap \mathcal{H}_{0}|}{|\hat{S}| \vee 1}\right].\nonumber%
\end{align}

An $\alpha$-level post-hoc $\mathrm{FDP}$ upper bound \cite{goeman2011multiple} is a function $V$ that verifies:
\begin{align}
\mathbb{P}\left(\forall S \subset \llbracket p \rrbracket, \mathrm{FDP}(S)  \leq V(S)/|S|\right) \geq 1-\alpha \,.\nonumber
\label{eq:FDPcontrol}
\end{align}
\textbf{Knockoffs.} The Knockoff filter is a variable selection technique introduced by \cite{barber2015controlling} and refined by \cite{candes2018panning} which controls the FDR. This procedure relies on building noisy copies of the original variables called Knockoff variables, that are designed to serve as controls for variable selection.
%

\begin{definition}[Model-X Knockoffs, \citealp{candes2018panning}]
\label{def:MXKnockoffs}
For the family of random variables $\mathbf{x} = (x_1, \ldots, x_p)$, Knockoffs are a new family of random variables $\mathbf{\tilde{x}} = (\tilde{x}_1, \ldots,\tilde{x}_p)$ satisfying:
\begin{enumerate}
\item for any $S \subset \llbracket p \rrbracket$, $(\mathbf{x}, \mathbf{\tilde{x}})_{swap(S)} \stackrel{d}{=} (\mathbf{x}, \mathbf{\tilde{x}})$
\item $\mathbf{\tilde{x}} \perp \mathbf{y} | \mathbf{x}$.
\end{enumerate}
\end{definition}
Once we have such variables at our disposal, we quantify their importance relative to the original ones. This is done by computing Knockoff statistics $\mathbf{W} = (W_1, \ldots, W_p)$ that are defined as follows.
\begin{definition}[Knockoff Statistic, \citealp{candes2018panning}]
  \label{def:KO-stat}
  A knockoff statistic $\mathbf{W} = (W_1, \ldots, W_p)$ is a measure of feature importance that satisfies:
  \begin{enumerate}
  \item $\mathbf{W}$ depends only on $\mathbf{X}, \mathbf{\tilde{X}}$ and $\mathbf{y}$:
  $\mathbf{W} = g(\mathbf{X}, \mathbf{\tilde{X}}, \mathbf{y})$.
  \item Swapping column $\mathbf{x}_j$ and its knockoff column $\mathbf{\tilde{x}}_j$ switches the sign of $W_j$: 
    \begin{equation*}
      W_j([\mathbf{X}, \mathbf{\tilde{X}}]_{swap(S)}, \mathbf{y}) =
      \left\{
          \begin{array}{ll}
            W_j([\mathbf{X}, \mathbf{\tilde{X}}], \mathbf{y}) \ \text{if } j \in S^{c} \\
            -W_j([\mathbf{X}, \mathbf{\tilde{X}}], \mathbf{y}) \ \text{if } j \in S . \\
          \end{array}
      \right.
    \end{equation*}
  \end{enumerate}
\end{definition}

The most commonly used Knockoff statistic is the Lasso-coefficient difference (LCD) \citep{weinstein2020power}. This statistic is obtained by fitting a Lasso estimator \citep{tibshirani1996regression} on $[\mathbf{X}, \mathbf{\tilde{X}}] \in \mathbb{R}^{n \times 2p}$, which yields $\widehat{\boldsymbol{\beta}} \in \mathbb{R}^{2p}$. Then, the Knockoff statistic can be computed using $\widehat{\boldsymbol{\beta}}$:
\begin{align}
\forall j  \in \llbracket p \rrbracket, \quad W_j = |\widehat{\beta}_j| - |\widehat{\beta}_{j+p}|.\nonumber
\end{align}
This coefficient summarizes the importance of the original $j^{th}$ variable relative to its own Knockoff: $W_j > 0$ indicates that the original variable is more more important for fitting $y$ than the Knockoff variable, meaning that the $j^{th}$ variable is likely relevant. Conversely, $W_j < 0$ indicates that the $j^{th}$ variable is probably irrelevant.
We thus wish to select variables corresponding to large and positive $W_j$. Formally, the rejection set $\hat{S}$ can be written $\hat{S} = \{j : W_j > T_q\},$ where $T_q$ is chosen to provably control the $\operatorname{FDR}$ at level $q$ \cite{candes2018panning}.

\textbf{Aggregation schemes.} Due to the randomness in the knockoff generation process, different variables may be selected for two different runs of the method, which is undesirable. To mitigate this, aggregation of multiple Knockoffs runs is needed. \citet{ren2022derandomized} introduced an aggregation scheme which relies defining Knockoffs $e$-values.
\begin{align}
e_{j}=
\frac{p}{1+|\left\{k: W_{k} \leq-T_q\right\}|}1_{\left\{W_j \geq T_q\right\}}.\nonumber
\end{align}
Such e-values can be averaged across $D$ draws and e-BH \citep{wang2022false} is performed for variable selection. Alternatively, \cite{nguyen2020aggregation} defines the following $\pi$-statistic, that quantifies the evidence against a variable:
\begin{equation}
\pi_{j}=\left\{\begin{array}{l}
\frac{1+|\left\{k: W_{k} \leq-W_{j}\right\}|}{p} \text { if } \quad W_{j} > 0\\
1 \quad \text { if } \quad W_{j} \leq 0.
\end{array}\right.
\label{eq:proportion}
\end{equation}

In \cite{nguyen2020aggregation} $\pi$ statistics are treated as $p$-values and aggregated using quantile aggregation \citep{meinshausen2009p}. However, they can only be considered $p$-values under restrictive assumptions that are hard to check.
In the next section, these statistics are used as a building block to reach FDP control. The KOPI framework does not require $\pi$ statistics to be valid $p$-values.

\section{Main contribution: FDP control for aggregated Knockoffs}
\label{sec:main-contr-fdp}
\subsection{Post hoc FDP control for $\pi$ statistics}
To obtain FDP control, we rely on Joint Error Rate control as introduced in \cite{blanchard2020post}.
For $k_{max} \in \llbracket p\rrbracket$, we define a \emph{threshold family} of size $k_{max}$ as a vector $\mathbf{t} = (t_j)_{j \in \llbracket k_{max} \rrbracket}$ such that $0 \leq t_1 \leq \dots \leq t_{k_{max}} \leq 1$.
\begin{definition}[Joint Error Rate, \citealp{blanchard2020post}]
\label{def:JER}
Denote by $\pi_{(j : \mathcal{H}_0)}$ the $j^{th}$ smallest value $\pi_{j}$ amongst all null hypotheses. The JER associated with $\mathbf{t} = (t_j)_{j \in \llbracket k_{max} \rrbracket}$ is:
\begin{align}
\label{eq:JERdef}
\operatorname{JER}(\mathbf{t})=\mathbb{P}\left(\exists j \in\llbracket k_{max} \wedge p_{0} \rrbracket: \pi_{\left(j:\mathcal{H}_{0}\right)}<t_{j}\right).
\end{align}

The threshold family $\mathbf{t}$ is said to control the $\operatorname{JER}$ at level $\alpha$ iff $\operatorname{JER}(\mathbf{t}) \leq \alpha$.
\end{definition}

An $\alpha$-level $\operatorname{FDP}$ upper bound can be derived from $\operatorname{JER}$ control via the following result:

\begin{restatable}[FDP control via JER control \citealp{blanchard2020post}]{proposition}{fdpbound}

\label{prop:bound}
If $\mathbf{t}$ is a threshold family of length $k_{max}$ that controls the $\operatorname{JER}$ at level $\alpha$, then, $V^{\mathbf{t}}(S) / |S|$ is an $\alpha$-level FDP upper bound, with:
\begin{align}
  V^{\mathbf{t}}(S)=\min _{1 \leq k \leq k_{max}} (k-1) + \sum_{i \in S} 1_{\left\{\pi_i > t_{k}\right\}}.
\label{eq:bound}
\end{align}
    
\end{restatable}

The proof of this result -- originally included in \cite{blanchard2020post} -- can be found in appendix \ref{sec:proofFDP} for self-containedness. In the remainder of this section, we show how to obtain JER control for $\pi$ statistics.

\subsection{Joint distribution of $\pi$ statistics under the null}
By Definition~\ref{def:JER}, JER$(\mathbf{t})$ of a given threshold family only depends on the joint null distribution of the $\pi$ statistics.
As for earlier FDR control \cite{barber2015controlling} or k-FWER control \cite{janson2016familywise} results, the key idea to obtain JER control for $\pi$ statistics is to prove that the relevant part of this distribution is in fact known, thanks to the properties of knockoff statistics.
%
We use the same notation as in \cite{janson2016familywise}.
Letting $Z_j = |\left\{k \in \llbracket p\rrbracket: W_{k} \leq-W_{j}\right\}|$ and $\chi_j = sign(W_j)$, the $\pi$ statistics $(\pi_j)_{j= \llbracket p\rrbracket}$ are given by:
\begin{align}
\pi_{j}=\frac{1 + Z_{j}}{p} 1_{\{\chi_{j} = 1\}} + 1_{\{\chi_{j} = -1\}}.\nonumber
\end{align}
For a given $\mathbf{W}$, let $\sigma(\mathbf{W})$ be a permutation of $\llbracket p\rrbracket$ that sorts $\mathbf{W}$ by decreasing modulus: $\sigma(\mathbf{W}) = (\sigma_1, \dots, \sigma_p)$ such that $|W_{\sigma_1}| \geq |W_{\sigma_2} |\cdots \geq |W_{\sigma_p}|$.
We start by proving that the $Z$ statistics can be expressed as a function of the vector of $\chi$ statistics:
\begin{lemma}
  \label{lem:Z-chi}
  For $k \in \llbracket p\rrbracket$ such that $\chi_{\sigma_k}=1$,
    $Z_{\sigma_k} = \sum_{j=1}^{k-1} 1_{\{\chi_{\sigma_j} = -1\}}.$
\end{lemma}
\begin{proof}[Proof of Lemma~\ref{lem:Z-chi}]
  We have
  \begin{align*}
    Z_j
    & = |\left\{k \in \llbracket p\rrbracket: W_{k} \leq-W_{j}\right\}| \\
    & = |\left\{k \in \llbracket p\rrbracket: W_{k} < 0 \textrm{ and } W_{k} \leq -W_{j}\right\}|
  \end{align*}
  If $W_{k} < 0$, then $W_{k} \leq -W_{j}$ is equivalent to $|W_{k}| \geq |W_{j}|$, which holds if and only if $k < j$ by the definition of $\sigma(\mathbf{W})$.%
  \iPN{should be `` if and only if $k \leq j$''?}
\end{proof}
Lemma~\ref{lem:Z-chi} implies that the distribution of order statistics of $\pi | \sigma(\mathbf{W})$ is entirely determined by that of $\chi | \sigma(\mathbf{W})$. To formalize this, we introduce $\pi^0$ statistics.
\begin{definition}[$\pi^0$ statistics]
  \label{def:pi0}
  Let $\chi^0=(\chi^0_j)_{1 \leq j \leq p}$ be a collection of $p$ i.i.d. Rademacher random variables, that is, for all $j$, $\mathbb{P}(\chi^0_j = 1) = \mathbb{P}(\chi^0_j = -1) = 1/2$.
  The associated $\pi^0$ statistics are defined for $j \in \llbracket p\rrbracket$ by 
\begin{align}
  \pi^0_{j}=\frac{1 + Z^0_{j}}{p} 1_{\{\chi^0_{j} = 1\}} + 1_{\{\chi^0_{j} = -1\}},
  \label{eq:def-pi0}
\text{ where }
Z^0_{j} = \sum_{k=1}^{j-1} 1_{\{\chi^0_{k} = -1\}}.
\end{align}
\end{definition}
\begin{theorem}
\label{thm:PIdistrib}
Let $\mathbf{t}$ be a threshold family of length $k_{max}$.
Then, for $\pi^0 = (\pi^0_j)_{j \in \llbracket p \rrbracket}$ as in \eqref{eq:def-pi0},
\begin{equation}
  \label{eq:JER-known-H0}
  \operatorname{JER}{\mathbf(\mathbf{t})} \leq
  \operatorname{JER}^0{\mathbf(\mathbf{t})} :=
  \mathbb{P}
  \left(
  \exists k \in \llbracket k_{max} \rrbracket: \pi^0_{(k)} < t_k 
  \right).
\end{equation}
\end{theorem}
\begin{proof}[Proof of Theorem~\ref{thm:PIdistrib}]
 Let $k \in \llbracket k_{max} \rrbracket$.
 Since $t_k \leq 1$, we have $\pi_{(k:\mathcal{H}_0)} < t_k$ if and only if $N_k \geq k$, where 
  \begin{align}
N_k =  \left| \left\{
      j \in \mathcal{H}_0, \chi_j=1 \textrm{ and } \frac{1 + Z_j}{p} < t_k
     \right\} \right|.\nonumber
  \end{align}
  With the notation of Definition~\ref{def:pi0}, we define the random variable
  \begin{align}
N^0_k =  \left| \left\{
      j \in \mathcal{H}_0, \chi^0_j=1 \textrm{ and } \frac{1 + Z^0_j}{p} < t_k
     \right\}\right|.\nonumber
  \end{align}
  If $\mathcal{H}_0 = \llbracket p\rrbracket$, then Lemma ~\ref{lem:Z-chi} implies that conditional on $\sigma(\mathbf{W})$, $N_k$ and $N^0_k$ have the same distribution.
  Indeed, the vectors $(W_j)_{j/ \chi_j = 1}$ and $(Z_j)_{j/ \chi_j = 1}$ have the same ordering, and conditional on $\sigma(\mathbf{W})$, $(\chi_j)_{j \in \mathcal{H}_0}$ are jointly independent and uniformly distributed on $\{-1,1\}$ (Lemma 2.1 in \citealp{janson2016familywise}; \citealp{barber2015controlling}).
  Using the same argument as in the proof of Lemma 3.1 in \citet{janson2016familywise}, in the case where $\mathcal{H}_0 \subsetneq \llbracket p\rrbracket$, false null $\chi_j$ will insert $-1$'s into the process on the nulls, implying that $N_k$ is stochastically dominated by $N^0_k$.
  Noting that $N^0_k \geq k$ if and only if $\pi^0_{(k)} < t_k$, we obtain that
  \begin{align*}
  \mathbb{P} \left(
    \exists k \in  \llbracket k_{max} \wedge p_{0} \rrbracket,
    \pi_{(k:\mathcal{H}_0)} < t_k
    | \sigma(\mathbf{W})
  \right)
  & \leq 
  \mathbb{P} \left(
    \exists k \in  \llbracket k_{max}  \wedge p_{0} \rrbracket,
    \pi^0_{(k)} < t_k
  \right) \\
    & \leq 
      \mathbb{P} \left(
      \exists k \in  \llbracket k_{max} \rrbracket,
      \pi^0_{(k)} < t_k
      \right).
  \end{align*}
  Taking the expectation with respect to $\sigma(\mathbf{W})$ yields the desired result.%
\end{proof}
%
Theorem~\ref{thm:PIdistrib} is related to Lemma 3.1 of \citet{janson2016familywise} and Lemma 3.1 of \citet{li2022simultaneous}, that rely on the sign-flip property of Knockoff statistics under the null \citep{barber2015controlling}.
The interest of Theorem~\ref{thm:PIdistrib} is that the upper bound $\operatorname{JER}^0{\mathbf(\mathbf{t})}$ only depends on the $\pi^0$ statistics and the threshold family $\mathbf{t}$, and not on the original data. Therefore, it can be estimated with arbitrary precision for any given $\mathbf{t}$ using Monte-Carlo simulation, as explained in the next section and described in Algorithm~\ref{alg:nullp} in Supp. Mat.
\subsection{Joint Error Rate control for $\pi$ statistics via calibration}
To approximate the $\operatorname{JER}$ upper bound derived in Theorem \ref{thm:PIdistrib}, we draw $B$ Monte-Carlo samples using Algorithm \ref{alg:nullp}. This yields a set of $B$ vectors of $\pi^0$ statistics denoted by $\pi^0_{b} \in \mathbb{R}^p$ for each $b \in \llbracket B \rrbracket$. This allows us to evaluate the empirical $\operatorname{JER}$, which estimates the upper bound of interest. 
\begin{definition}[Empirical JER]
For $B$ vectors of $\pi^0$ statistics and a threshold family $\mathbf{t}$, the empirical $\operatorname{JER}$ is defined as:
\begin{align}
\label{eq:empiricalJER}
\widehat{\operatorname{JER}}^0_B(\mathbf{t}) = \frac{1}{B} \sum_{b=1}^{B} 1\left\{\exists k \in \llbracket k_{max} \rrbracket: \pi^0_{b(k)}<t_{k}\right\},
\end{align}
where for each $b \in \llbracket B \rrbracket$, $\pi^0_{b(1)} \leq \dots \leq \pi^0_{b(p)}$.
\end{definition}
Since $\widehat{\operatorname{JER}}^0_B(\mathbf{t})$ can be made arbitrarily close (by choosing $B$ large enough) to $\widehat{\operatorname{JER}}^0(\mathbf{t})$ for any given threshold family $\mathbf{t}$, it remains to choose  $\mathbf{t}$ such that $\widehat{\operatorname{JER}}^0(\mathbf{t}) \leq \alpha$ in order to ensure $\operatorname{JER}$ control. 
To this end, we consider a sorted set of candidate threshold families called a \emph{template}:
\begin{definition}[Template \cite{blanchard2020post}]
A template is a component-wise non-decreasing function $\textbf{T} : [0, 1] \mapsto \mathbb{R}^{p}$ that maps a parameter $\lambda \in [0, 1]$ to a threshold family $\mathbf{T}(\lambda) \in \mathbb{R}^{p}$.

This definition is naturally extended to the case of templates containing a finite number of threshold families. The template corresponding to $B'$ threshold families is then denoted by
$\left(\textbf{T}\left(b'/B'\right)\right)_{b' \in \llbracket B' \rrbracket}$.
\end{definition}
Once a template is specified, the \textit{calibration} procedure \citep{blanchard2020post} can be performed; this consists in finding the least conservative threshold family $\mathbf{t}$ amongst the template that controls the empirical $\operatorname{JER}$ at level $\alpha$. 
Formally, we consider the threshold family defined $\mathbf{t}^B_\alpha = \mathbf{T}(\lambda_B(\alpha))$, where
\begin{align}
\lambda_B(\alpha) = \frac{1}{B'}\max \left\{ b' \in \llbracket B' \rrbracket \quad s.t. \quad  \widehat{\operatorname{JER}}^0_{B}\left(\mathbf{T}\left(\frac{b'}{B'}\right)\right) \leq \alpha \right\}.\nonumber
\end{align}

As observed by \citet{blain2022notip}, optimal power is reached when the candidate families match the shape of the distribution of the null statistics. 
We define a template based on the distribution of the $\pi^0$ statistics appearing in Theorem \ref{thm:PIdistrib}. In practice, we draw $B'$ samples from this distribution independently from the $B$ Monte Carlo samples to avoid circularity biases. 
Since a template has to be component-wise non-decreasing, i.e. the set of candidate threshold families has to be sorted, we extract empirical quantiles from these $B'$ sorted vectors. 
This yields a template $\mathbf{T}^0$ composed of $B'$ candidate curves that match quantiles of the distribution of $\pi^0$ statistics. The $\frac{b'}{B'}$-quantile curve defines the threshold family $\mathbf{T}^0\left(b'/B'\right)$.
We obtain the following result:
\begin{restatable}[JER control for $\pi$-statistics]{theorem}{JERPi}
\label{thm:JERpi1}
Consider the threshold family defined by $\mathbf{t}^B_\alpha = \mathbf{T}^0(\lambda_B(\alpha))$.
Then, as  $B \to +\infty$, 
\begin{align*}
\operatorname{JER}(\mathbf{t}^B_\alpha) \leq \alpha + O_P(1/\sqrt{B}).
\end{align*}
\end{restatable}

The number $B$ of Monte-Carlo samples in Theorem \ref{thm:JERpi1} can be chosen arbitrarily large to obtain $\operatorname{JER}$ control, leading to valid $\operatorname{FDP}$ bounds via Equation \ref{eq:bound}. This result is proved in Appendix \ref{sec:proofthm2}.

\subsection{False Discovery Proportion control for aggregated Knockoffs}
In the previous section we have seen how to reach FDP control via Knockoffs. As explained above, aggregation is needed to mitigate the randomness of the Knockoff generation process. Therefore, we aim to extend the previous result to the case of aggregated Knockoffs. Let us first define aggregation: 

\begin{definition}
    For $D$ draws of Knockoffs, an aggregation procedure is a function $f: \mathbb{R}^D \mapsto \mathbb{R}$ that maps a vector of $(\pi^d)_{d\in \llbracket D \rrbracket}$ statistics to an aggregated statistic $\overline{\pi}$.
\end{definition}

In practice, since we have $p$ variables, aggregation is performed for each variable, i.e.:
    \begin{align}
    \forall j \in \llbracket p \rrbracket, \quad f(\pi_j^{1}, \ldots,\pi_j^{D}) = \overline{\pi_j}.\nonumber
    \end{align}
Then, inference is performed on the vector of aggregated statistics $(\overline{\pi}_1,\ldots, \overline{\pi}_p)$.

For a fixed aggregation scheme $f$, we can naturally extend the calibration procedure of the preceding section. Instead of drawing a single $B \times p$ matrix of $\pi^0$ statistics containing $\pi^0_b \in \mathbb{R}^p$ for each $b \in \llbracket B \rrbracket$, we draw $D$ such matrices. Given $d \in \llbracket D \rrbracket$, each matrix contains $\pi^{0,d}_{b} \in \mathbb{R}^p$ for each $\llbracket B \rrbracket$.

Then, for each $b \in \llbracket B \rrbracket $, we perform aggregation:
%
$
\overline{\pi}^{0}_{b} = f\left((\pi^{0,d}_{b})_{d \in \llbracket D \rrbracket}\right).
$
%
%
The $\operatorname{JER}$ in the aggregated case is defined as:
\begin{align}
\overline{\operatorname{JER}}(\mathbf{t})=\mathbb{P}\left(\exists j \in\llbracket k_{max} \wedge p_{0} \rrbracket: \overline{\pi}_{\left(j:\mathcal{H}_{0}\right)}<t_{j}\right).
\nonumber\end{align}%
We obtain the aggregated template following the same procedure, i.e. drawing $D$ templates and aggregating them. For each $b' \in \llbracket B'\rrbracket$, the aggregated threshold family is written:
\begin{align}\overline{\mathbf{T}}\left(\frac{b'}{B'}\right) = f\left(\left(\mathbf{T}^d\left(\frac{b'}{B'}\right)\right)_{d \in \llbracket D \rrbracket}\right).\nonumber\end{align}
We can then write the empirical $\operatorname{JER}$ in the aggregated case as:
\begin{align}
& \widehat{\overline{\operatorname{JER}}}\left(\overline{\mathbf{T}}\left(\frac{b'}{B'}\right)\right) = \frac{1}{B} \sum_{b=1}^{B} 1\left\{
\exists j \in \llbracket k_{max} \rrbracket: \overline{\pi}^0_{b(j)}<\overline{\mathbf{T}}_j\left(\frac{b'}{B'}\right)
\right\}. \nonumber
\end{align}
Calibration can be performed in the same way as in the non-aggregated case. Note that we perform calibration \emph{after} aggregating; therefore, $\operatorname{JER}$ control is ensured directly on aggregated statistics and is not a result of aggregating $\operatorname{JER}$ controlling families. Importantly, this approach holds without additional assumptions on the aggregation scheme $f$. We consider the threshold family $\overline{\mathbf{t}}^B_\alpha = \overline{\mathbf{T}}(\lambda_B(\alpha))$, where
\begin{align}
\lambda_B(\alpha) = \frac{1}{B'}\max \left\{ b' \in \llbracket B' \rrbracket \quad s.t. \quad  \widehat{\overline{\operatorname{JER}}}^0_{B}\left(\overline{\mathbf{T}}\left(\frac{b'}{B'}\right)\right) \leq \alpha \right\}.\nonumber
\end{align}

With $\overline{\mathbf{T}}^0$ a template composed of $B'$ candidate curves that match quantiles of the distribution of $\overline{\pi}^0$ statistics, we obtain the following result:
\begin{theorem}[JER control for aggregated $\pi$-statistics]
\label{thm:JERpi2}
Consider the threshold family defined by $\overline{\mathbf{t}}^B_\alpha = \overline{\mathbf{T}}^0(\lambda_B(\alpha))$.
Then, as  $B \to +\infty$, 
\begin{align*}
\overline{\operatorname{JER}}(\overline{\mathbf{t}}^B_\alpha) \leq \alpha + O_P(1/\sqrt{B}).
\end{align*}
\end{theorem}
\begin{proof}
The proof is identical to that of Theorem \ref{thm:JERpi1} using the empirical aggregated $\operatorname{JER}$. 
\end{proof}
The calibrated aggregated threshold family yields valid $\operatorname{FDP}$ upper bounds via Proposition \ref{prop:bound}. 
The proposed \textbf{KOPI} (Knockoffs - $\pi$) method therefore achieves FDP control on aggregated Knockoffs.

\section{Experiments}

\textbf{Methods considered.} In our implementation of KOPI, we rely on the harmonic mean \citep{wilson2019harmonic} as the aggregation scheme $f$. Additionally, we set $k_{max} = \lfloor p / 50 \rfloor $ following the approach of \cite{blain2022notip}. We also consider both state-of-the-art Knockoffs aggregation schemes: AKO (Aggregation of Multiple Knockoffs, \citealp{nguyen2020aggregation}) and e-values based aggregation \citep{ren2022derandomized}. Additionally, we consider Vanilla Knockoffs, i.e. \cite{candes2018panning} and $\operatorname{FDP}$ control via Closed Testing \citep{li2022simultaneous}.
In simulated data experiments, we generate Knockoffs assuming a Gaussian distribution for  $\mathbf{X}$, with all variables centered.
For methods that support aggregation, we use $D = 50$ Knockoff draws.

\subsection{Simulated data}
\textbf{Setup.} At each simulation run, we generate Gaussian data $\mathbf{X} \in \mathbb{R}^{n \times p}$ with a Toeplitz correlation matrix corresponding to a first-order auto-regressive model with parameter $\rho$, i.e. $\mathbf{\Sigma}_{i, j} = \rho^{|i-j|}$.

Then, we draw the true support $\boldsymbol{\beta}^* \in \{0, 1\}^p$. The number of non-null coefficients of $\boldsymbol{\beta}^*$ is controlled by the sparsity parameter $s_p$, i.e. $s_p = \lVert \boldsymbol{\beta}^* \rVert_0/p$. The target variable $\mathbf{y}$ is built
using a linear model:
\begin{align}\mathbf{y} = \mathbf{X}\boldsymbol{\beta}^* + \sigma\boldsymbol{\epsilon},\nonumber\end{align}
with $\sigma$ controlling the amplitude of the noise: $\sigma = \lVert \mathbf{X} \boldsymbol{\beta}^* \rVert_2 /
(\text{SNR}\lVert\boldsymbol{\epsilon}\rVert_2)$, SNR being the signal-to-noise ratio.
We choose the central setting $n = 500, p = 500, \rho = 0.5, s_p = 0.1, \text{SNR} = 2$. For each parameter, we explore a range of possible values to benchmark the methods across varied settings.

\begin{figure}[t]
\centering
\includegraphics[width=1\linewidth]{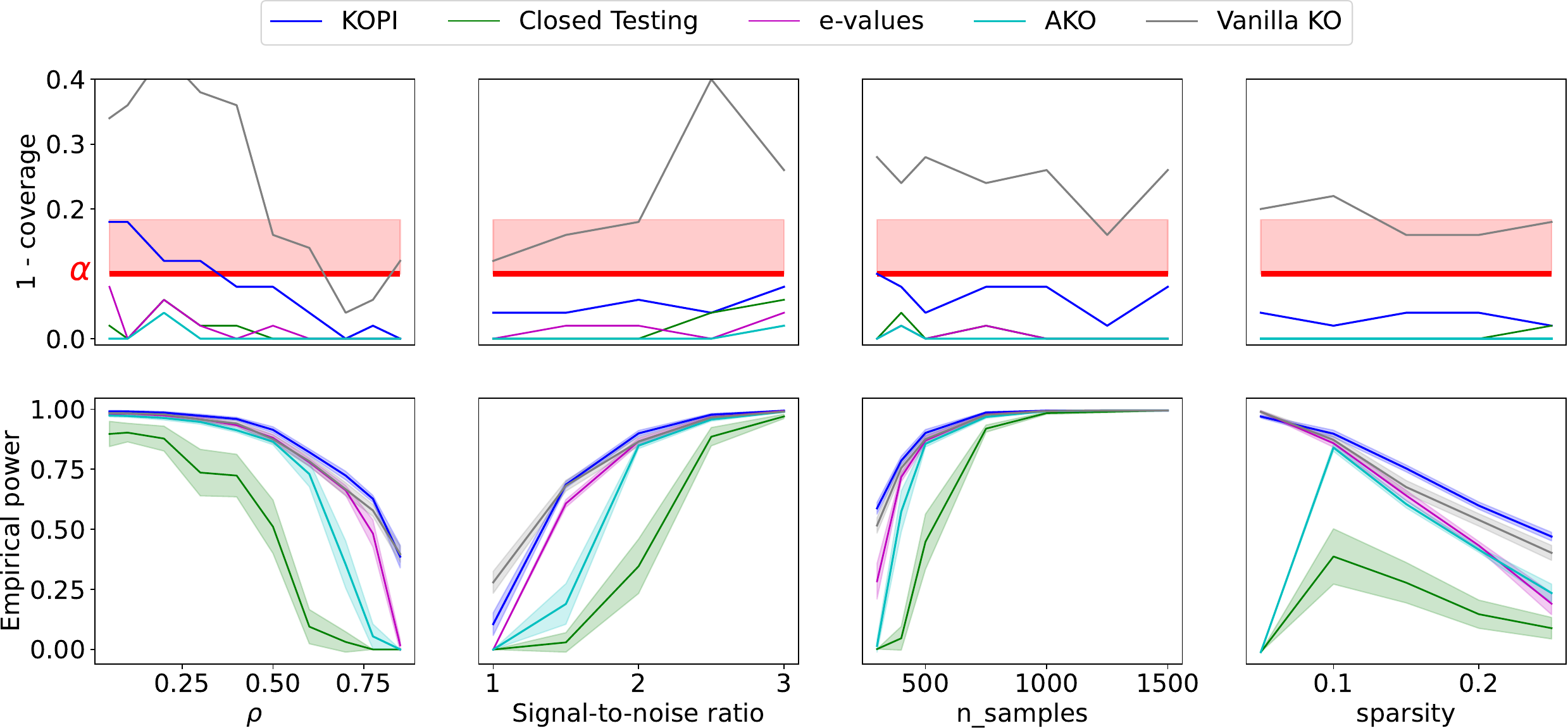}
\caption{\textbf{FDP bound coverage at level $\alpha$ and empirical Power for 50 simulation runs and five different methods:} Vanilla Knockoffs, aggregated Knockoffs using e-values, aggregated Knockoffs using quantile-aggregation, KOPI and Knockoff inference via Closed Testing. We use $D=50$ Knockoffs draws and the following simulation settings: $\alpha = 0.1, q = 0.1, p = 500$. Each column represents a varying parameter with the first row displaying FDP coverage and the second row displaying power. The red line and associated error bands represent the acceptable limits for FDP bound coverage. KOPI consistently outperforms all other methods while retaining FDP control.}
\vskip-1em
\label{fig:sim1}
\end{figure}

\looseness=-1
To select variables using $\operatorname{FDP}$ upper bounds, we retain the largest possible set of variables $S$ such that $V(S) \leq q |S|$ (Algorithm~\ref{alg:inference}).
For each of the $N$ simulations and each method, we compute the empirical FDP and True Positive Proportion (TPP): \begin{align}\widehat{FDP}(S) = \frac{|S\cap \mathcal{H}_0|}{|S|} 
\quad \text{and} \quad
\widehat{TPP}(S) = \frac{|S\cap \mathcal{H}_1|}{|\mathcal{H}_1|}.\nonumber\end{align}
%
If the $\operatorname{FDP}$ is controlled at level $\alpha$, $|\{k \in \llbracket N \rrbracket : \widehat{FDP}(S_k) > q \}| \sim \mathcal{B}(N, \alpha)$.
Then, we can compute error bands on the $\alpha$-level using $\operatorname{std}\left(\mathcal{B}(N, \alpha)/N\right) = \sqrt{\alpha(1-\alpha)/N}$.
The second row of Fig. \ref{fig:sim1} represents the empirical power achieved by each method, which corresponds to the average of TPPs defined above for $N$ runs i.e. $\text{Power} = \sum_{k=1}^{N}\widehat{TPP}(S_k)/N.$
Fig. \ref{fig:sim1} shows that across all different settings, KOPI retains FDP control. We can also see that FDR control does not imply FDP control, as Vanilla Knockoffs are consistently outside of FDP bound coverage intervals. However, the two existing aggregation schemes (AKO and e-values) that formally guarantee FDR control are generally conservative and achieve FDP control empirically. This is consistent with the findings of \cite{ren2022derandomized}. The Closed Testing procedure of \cite{li2022simultaneous} achieves FDP control as announced but suffers from a lack of power.

%
Interestingly, KOPI achieves FDP control while offering power gains compared to FDR-controlling Knockoffs aggregation methods. Yet FDP control is a much stronger guarantee than FDR control, as discussed previously. These gains are especially noticeable in challenging inference settings where most methods exhibit a clear decrease in power or even catastrophic behavior (i.e. zero power).

Moreover, Fig. \ref{fig:sim2} (in appendix) shows that when using $q = 0.05$ rather than $q = 0.1$ as in Fig. \ref{fig:sim1}, the robustness of KOPI with regards to difficult inference settings is even more salient. More precisely, for $q = 0.05$, AKO and Closed Testing are always powerless. E-values aggregation yields good power in easier settings such as $\rho \leq 0.6$, $\text{SNR} \geq 2.5$ or $n > 750$ but exhibits catastrophic behavior in harder settings. Overall, apart from KOPI, only Vanilla Knockoffs exhibit non-zero power, but this method fails to control the FDP as it is intended to control FDR. KOPI preserves FDP control in all settings while yielding superior power compared to all other methods.

\subsection{Brain data application}

The goal of human brain mapping is to associate cognitive tasks with relevant brain regions. This problem is tackled using functional Magnetic Resonance Imaging (fMRI), which consists in recording the blood oxygenation level dependent signal via an MRI scanner. 
The importance of conditional inference for this problem has been outlined in \cite{weichwald2015causal}.
We use the Human Connectome Project (HCP900) dataset that contains brain images of healthy young adults performing different tasks while inside an MRI scanner. Details about this dataset and empirical results can be found in Appendix \ref{sec:HCPappendix}.
%
%

While these results demonstrate the face validity of the approach, FDP control and power cannot be evaluated. Therefore, following \cite{Nguyen2022CRT}, we consider an additional experiment that consists in using semi-simulated data.
We consider a first fMRI dataset $(\mathbf{X}_1, \mathbf{y}_1)$ on which we perform inference using a Lasso estimator; this yields $\boldsymbol{\beta}_1^* \in \mathbb{R}^{p}$ that we will use as our ground truth. Then, we consider a separate fMRI dataset $(\mathbf{X}_2, \mathbf{y}_2)$ for data generation. 
The point of using a separate dataset is to avoid circularity between the ground truth definition and the inference procedure. 
Concretely, we discard the original response vector $\mathbf{y}_2$ for this dataset and build a simulated response $\mathbf{y}_2^{sim}$ using a linear model, with the same notation as previously (we set $\sigma$ so that $SNR = 4$):
%
$\mathbf{y}_2^{sim} = \mathbf{X}_2\boldsymbol{\beta}_1^* + \sigma\boldsymbol{\epsilon}.$

\begin{figure}[t]
    \centering
    \subfloat{{\includegraphics[width=0.45\linewidth]{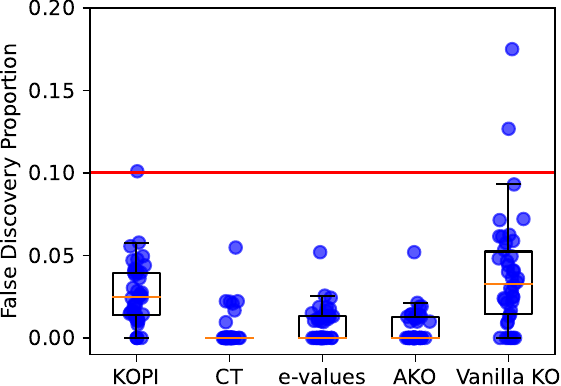} }}%
    \qquad
    \subfloat{{\includegraphics[width=0.45\linewidth]{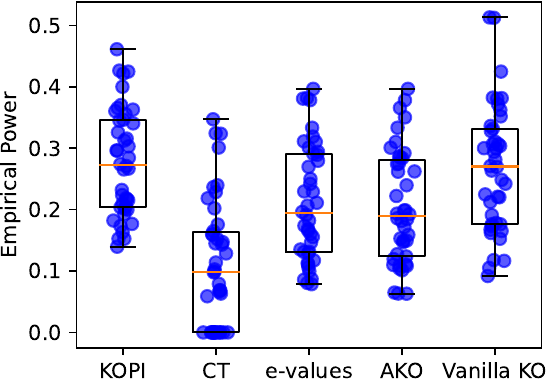} }}%
    \caption{\textbf{Empirical FDP and power on semi-simulated data for 42 contrast pairs.} We use 7 HCP contrasts C0: "Motor Hand", C1: "Motor Foot", C2: "Gambling", C3: "Relational", C4: "Emotion", C5: "Social", C6: "Working Memory". We consider all 42 possible train/test pairs: the train contrast is used to obtain a ground truth, while the test contrast is used to generate the response. Inference is performed using the 5 methods considered in the paper and the empirical FDP is reported in the left box plot, while power is reported in the right box plot.
Notice (right figure) that KOPI yields superior power compared to all other Knockoffs-based methods while controlling the FDP (left Fig.).
}%
    \label{fig:semisim}%
\end{figure}

Then, inference is performed using Knockoffs-based methods on $(\mathbf{X}_2, \mathbf{y}_2^{sim})$. Since we consider $\boldsymbol{\beta}_1^*$ as the ground truth, the FDP and TPP can be computed for each method. 
As can be seen in Fig. \ref{fig:semisim}, KOPI is the most powerful method among those that control the FDP.

\subsection{Genomic data application}

In addition to the brain data application, we compared KOPI to other Knockoffs-based methods on gene-expression data \cite{bourgon2010independent} containing $79$ samples and $90$ genes. KOPI yields a non trivial selection for all runs, with 3 genes selected in $100\%$ of all $50$ runs of the experiment. Across all runs, only $8$ different genes are selected by KOPI. Vanilla Knockoffs select $24$ different genes across all runs and no gene exceeds a selection frequency of $70\%$. All other methods are powerless in all runs. Details and results of this experiment can be found in Appendix \ref{sec:GenomicsExpe}.

\section{Discussion}
%
In this paper, we have proposed a novel method that reaches FDP control on aggregated Knockoffs. It combines the benefits of aggregation, i.e. improving the stability of the inference, in addition to providing a probabilistic control of the FDP, rather than controlling only its expectation, the FDR.

Simulation results support that KOPI indeed controls the FDP. Furthermore, while FDP control is a stricter guarantee than FDR control, KOPI actually offers power gains compared to state-of-the-art aggregation-based Knockoffs methods.
%
This sensitivity gain is a direct benefit from the JER approach and its adaptivity to arbitrary aggregation schemes. While the latter has been formulated and used so far in mass univariate settings \cite{blain2022notip}, the present work presents a first use of this approach in the context of multiple regression. 
%
Moreover, KOPI does not require any assumption on the data at hand or on the law of Knockoff statistics under the null.

The computation time of the proposed approach is comparable to existing aggregation schemes for Knockoffs: sampling $\pi$ statistics under the null using Algorithm \ref{alg:nullp} can be done once and for all for a given value of $p$. JER estimation via Algorithm \ref{alg:estimateJER} and calibration can be performed via binary search of complexity $\mathcal{O}(log(B'))$. Finding the rejection set $\hat{S}$ after performing calibration is done in linear time via \cite{enjalbert-courrech:powerful}.
In practice, the computation time is the same as for classical knockoff aggregation \citep{ren2021derandomizing} and is in minutes for the brain imaging datasets considered. Avenues for future work include a theoretical analysis of the False Negative Proportion (FNP) \cite{genovese2002operating} of KOPI and developing a step-down version of the method to further improve power.

We provide a Python package containing the code for KOPI available at \url{https://github.com/alexblnn/KOPI}.

\newpage

\section{Acknowledgments and disclosure of funding}

This project was funded by a UDOPIA PhD grant from Universit\'e Paris-Saclay and also supported by the FastBig ANR project (ANR-17-CE23-0011), the KARAIB AI chair (ANR-20-CHIA-0025-01), the H2020 Research Infrastructures Grant EBRAIN-Health 101058516 and the SansSouci ANR project (ANR-16-CE40-0019). 
The authors thank Binh Nguyen for his precious help on the code base and Samuel Davenport for useful discussions about this work.

\bibliographystyle{apalike}
\bibliography{bib}

\begin{thebibliography}{}

\bibitem[Barber and Cand{\`e}s, 2015]{barber2015controlling}
Barber, R.~F. and Cand{\`e}s, E.~J. (2015).
\newblock Controlling the false discovery rate via knockoffs.
\newblock {\em The Annals of Statistics}, 43(5):2055--2085.

\bibitem[Benjamini and Hochberg, 1995]{benjamini1995controlling}
Benjamini, Y. and Hochberg, Y. (1995).
\newblock Controlling the false discovery rate: a practical and powerful
  approach to multiple testing.
\newblock {\em Journal of the Royal statistical society: series B
  (Methodological)}, 57(1):289--300.

\bibitem[Blain et~al., 2022]{blain2022notip}
Blain, A., Thirion, B., and Neuvial, P. (2022).
\newblock Notip: Non-parametric true discovery proportion control for brain
  imaging.
\newblock {\em NeuroImage}, 260:119492.

\bibitem[Blanchard et~al., 2020]{blanchard2020post}
Blanchard, G., Neuvial, P., Roquain, E., et~al. (2020).
\newblock Post hoc confidence bounds on false positives using reference
  families.
\newblock {\em Annals of Statistics}, 48(3):1281--1303.

\bibitem[Bourgon et~al., 2010]{bourgon2010independent}
Bourgon, R., Gentleman, R., and Huber, W. (2010).
\newblock Independent filtering increases detection power for high-throughput
  experiments.
\newblock {\em Proceedings of the National Academy of Sciences},
  107(21):9546--9551.

\bibitem[Cand{\`e}s et~al., 2018]{candes2018panning}
Cand{\`e}s, E., Fan, Y., Janson, L., and Lv, J. (2018).
\newblock Panning for gold:‘model-x’knockoffs for high dimensional
  controlled variable selection.
\newblock {\em Journal of the Royal Statistical Society: Series B (Statistical
  Methodology)}, 80(3):551--577.

\bibitem[Enjalbert-Courrech and Neuvial, 2022]{enjalbert-courrech:powerful}
Enjalbert-Courrech, N. and Neuvial, P. (2022).
\newblock {Powerful and interpretable control of false discoveries in two-group
  differential expression studies}.
\newblock {\em Bioinformatics}, 38(23):5214--5221.

\bibitem[Genovese and Wasserman, 2002]{genovese2002operating}
Genovese, C. and Wasserman, L. (2002).
\newblock Operating characteristics and extensions of the false discovery rate
  procedure.
\newblock {\em Journal of the Royal Statistical Society Series B: Statistical
  Methodology}, 64(3):499--517.

\bibitem[Gimenez and Zou, 2019]{gimenez2019improving}
Gimenez, J.~R. and Zou, J. (2019).
\newblock Improving the stability of the knockoff procedure: Multiple
  simultaneous knockoffs and entropy maximization.
\newblock In {\em The 22nd International Conference on Artificial Intelligence
  and Statistics}, pages 2184--2192. PMLR.

\bibitem[Goeman and Solari, 2011]{goeman2011multiple}
Goeman, J.~J. and Solari, A. (2011).
\newblock Multiple testing for exploratory research.
\newblock {\em Statistical Science}, 26(4):584--597.

\bibitem[Janson and Su, 2016]{janson2016familywise}
Janson, L. and Su, W. (2016).
\newblock Familywise error rate control via knockoffs.
\newblock {\em Electronic Journal of Statistics}, 10(1):960--975.

\bibitem[Korn et~al., 2004]{korn2004controlling}
Korn, E.~L., Troendle, J.~F., McShane, L.~M., and Simon, R. (2004).
\newblock Controlling the number of false discoveries: application to
  high-dimensional genomic data.
\newblock {\em Journal of Statistical Planning and Inference}, 124(2):379--398.

\bibitem[Li et~al., 2022]{li2022simultaneous}
Li, J., Maathuis, M.~H., and Goeman, J.~J. (2022).
\newblock Simultaneous false discovery proportion bounds via knockoffs and
  closed testing.
\newblock {\em arXiv preprint arXiv:2212.12822}.

\bibitem[Luo et~al., 2022]{luo2022improving}
Luo, Y., Fithian, W., and Lei, L. (2022).
\newblock Improving knockoffs with conditional calibration.

\bibitem[Meinshausen et~al., 2009]{meinshausen2009p}
Meinshausen, N., Meier, L., and B{\"u}hlmann, P. (2009).
\newblock P-values for high-dimensional regression.
\newblock {\em Journal of the American Statistical Association},
  104(488):1671--1681.

\bibitem[Nguyen et~al., 2022]{Nguyen2022CRT}
Nguyen, B.~T., Thirion, B., and Arlot, S. (2022).
\newblock {A Conditional Randomization Test for Sparse Logistic Regression in
  High-Dimension}.
\newblock In {\em {NeurIPS 2022}}, volume~35 of {\em Advances in Neural
  Information Processing Systems}, New Orleans, United States.

\bibitem[Nguyen et~al., 2020]{nguyen2020aggregation}
Nguyen, T.-B., Chevalier, J.-A., Thirion, B., and Arlot, S. (2020).
\newblock Aggregation of multiple knockoffs.
\newblock In {\em International Conference on Machine Learning}, pages
  7283--7293. PMLR.

\bibitem[Ren and Barber, 2022]{ren2022derandomized}
Ren, Z. and Barber, R.~F. (2022).
\newblock Derandomized knockoffs: leveraging e-values for false discovery rate
  control.
\newblock {\em arXiv preprint arXiv:2205.15461}.

\bibitem[Ren et~al., 2021]{ren2021derandomizing}
Ren, Z., Wei, Y., and Cand{\`e}s, E. (2021).
\newblock Derandomizing knockoffs.
\newblock {\em Journal of the American Statistical Association}, pages 1--11.

\bibitem[Stiglic and Kokol, 2010]{stiglic2010stability}
Stiglic, G. and Kokol, P. (2010).
\newblock Stability of ranked gene lists in large microarray analysis studies.
\newblock {\em Journal of biomedicine and biotechnology}, 2010.

\bibitem[Thirion et~al., 2014]{thirion:ward}
Thirion, B., Varoquaux, G., Dohmatob, E., and Poline, J.-B. (2014).
\newblock {Which fMRI clustering gives good brain parcellations?}
\newblock {\em {Frontiers in Neuroscience}}, 8(167):13.

\bibitem[Tibshirani, 1996]{tibshirani1996regression}
Tibshirani, R. (1996).
\newblock Regression shrinkage and selection via the lasso.
\newblock {\em Journal of the Royal Statistical Society: Series B
  (Methodological)}, 58(1):267--288.

\bibitem[{Van Essen} et~al., 2012]{van2012}
{Van Essen}, D.~C., Ugurbil, K., Auerbach, E.~J., Barch, D.~M., Behrens, T.
  E.~J., Bucholz, R., Chang, A., Chen, L., Corbetta, M., Curtiss, S.~W., {Della
  Penna}, S., Feinberg, D.~A., Glasser, M.~F., Harel, N., Heath, A.~C.,
  Larson{-}Prior, L.~J., Marcus, D.~S., Michalareas, G., Moeller, S.,
  Oostenveld, R., Petersen, S.~E., Prior, F.~W., Schlaggar, B.~L., Smith,
  S.~M., Snyder, A.~Z., Xu, J., and Yacoub, E. (2012).
\newblock The {Human Connectome Project}: a data acquisition perspective.
\newblock {\em Neuroimage}, 62(4):2222--2231.

\bibitem[Vovk and Wang, 2021]{vovk2021values}
Vovk, V. and Wang, R. (2021).
\newblock E-values: Calibration, combination and applications.
\newblock {\em The Annals of Statistics}, 49(3):1736--1754.

\bibitem[Wang and Ramdas, 2022]{wang2022false}
Wang, R. and Ramdas, A. (2022).
\newblock False discovery rate control with e-values.
\newblock {\em Journal of the Royal Statistical Society Series B: Statistical
  Methodology}, 84(3):822--852.

\bibitem[Weichwald et~al., 2015]{weichwald2015causal}
Weichwald, S., Meyer, T., {\"O}zdenizci, O., Sch{\"o}lkopf, B., Ball, T., and
  Grosse-Wentrup, M. (2015).
\newblock Causal interpretation rules for encoding and decoding models in
  neuroimaging.
\newblock {\em NeuroImage}, 110:48--59.

\bibitem[Weinstein et~al., 2020]{weinstein2020power}
Weinstein, A., Su, W.~J., Bogdan, M., Barber, R.~F., and Cand{\`e}s, E.~J.
  (2020).
\newblock A power analysis for knockoffs with the lasso coefficient-difference
  statistic.
\newblock {\em arXiv preprint arXiv:2007.15346}.

\bibitem[Wilson, 2019]{wilson2019harmonic}
Wilson, D.~J. (2019).
\newblock The harmonic mean p-value for combining dependent tests.
\newblock {\em Proceedings of the National Academy of Sciences},
  116(4):1195--1200.

\end{thebibliography}

\newpage
\appendix
\begin{singlespace}
    \setstretch{2}
   \listofappendices 
\end{singlespace}
\newpage

\newappendix{Proofs}
\newsubappendix{Proof of FDP control via JER control}
\label{sec:proofFDP}
For self-containedness we provide a Proof of Proposition \ref{prop:bound} adapted from \cite{blanchard2020post}:

\fdpbound*
 
\begin{proof}
Denote by $R_k = \{j: \pi_j \leq t_k\}$. Then for any set $S$:
\begin{align*}
\left|S \cap {H}_{0}\right| 
&=\left|S \cap \overline{R_{k}}\cap {H}_{0}\right|+\left|S \cap R_{k} \cap {H}_{0}\right| \\ 
& \leq\left|S \cap \overline{R_{k}}\right|+\left|R_{k} \cap {H}_{0}\right| \\ 
& = \sum_{i \in S} 1_{\left\{\pi_{i}(X) > t_k\right\}} +\left|R_{k} \cap {H}_{0}\right| \\ 
& \leq \sum_{i \in S} 1_{\left\{\pi_{i}(X) > t_k\right\}} + k-1 \\
& =: V_k^{\mathbf{t}}(S) \, ,
\end{align*}
where the last inequality holds with probability at least $1 - \alpha$ by \eqref{eq:JERdef}. Since \eqref{eq:JERdef} holds simultaneously for all $k$, the minimum over $k$ of all $V_k(S)$ is an $\alpha$-level upper bound on the false positives in $S$ and therefore $V^{\mathbf{t}}(S) / |S|$ is itself an $\alpha$-level FDP upper bound.
\end{proof}

\newsubappendix{Proof of Theorem \ref{thm:JERpi1}}
\label{sec:proofthm2}
\begin{lemma}
  \label{lem:JERpi1}
  For any threshold family $\mathbf{t}$, we have
  \begin{align*}
    \operatorname{JER}^0{\mathbf(\mathbf{t})} - \widehat{\operatorname{JER}}^0_B(\mathbf{t}) = O_P(1/\sqrt{B})
  \end{align*}
\end{lemma}
\begin{proof}[Proof of Lemma~\ref{lem:JERpi1}]
Let $Z_B(\mathbf{t}) = \sqrt{B} \left( \operatorname{JER}^0{\mathbf(\mathbf{t})} - \widehat{\operatorname{JER}}^0_B(\mathbf{t}) \right)$.
By the Central Limit Theorem, we have
\begin{align*}
    Z_B(\mathbf{t}) \xrightarrow[B \rightarrow \infty]{d} Z(\mathbf{t}),
\end{align*}
where $Z(\mathbf{t})$ is a centered Gaussian random variable with variance 
$\sigma^2(\mathbf{t}) = \operatorname{JER}^0{\mathbf(\mathbf{t})}(1 - \operatorname{JER}^0{\mathbf(\mathbf{t})})$.
As such, for any $M > 0$, we have 
\begin{align*}
  \mathbb{P} \left( \left| Z_B(\mathbf{t}) \right| \geq M \right)
  \xrightarrow[B \rightarrow \infty]{}
  \mathbb{P} \left( \left|  Z(\mathbf{t}) \right| \geq M \right)
  .
\end{align*}
Since $\operatorname{JER}^0{\mathbf(\mathbf{t})} \leq 1$, we have $\sigma^2(\mathbf{t}) \leq \nicefrac{1}{4}$ for any $\mathbf{t}$, so that $Z(\mathbf{t})$ is stochastically dominated by $\mathcal{N}\left(0, \nicefrac{1}{4}\right)$, which does not depend on the threshold family $\mathbf{t}$.
As such, we have $\mathbb{P} \left( \left|  Z(\mathbf{t}) \right| \geq M \right)  =
2 \mathbb{P} \left(  Z(\mathbf{t}) \geq M \right) \leq 2 \overline{\Phi}(2M)$, where $ \overline{\Phi}$ denotes the tail function of the standard normal distribution. 
Since $\overline{\Phi}(x)$ tends to $0$ as $x \to +\infty$, we have proved that
$Z_B(\mathbf{t}) = O_P(1)$.
\end{proof}

\JERPi*

\begin{proof}

We treat the case where $\mathbf{t}^B_\alpha$ is well defined for all $B$, i.e. that there exists a threshold family amongst $\mathbf{T}^0$ controls the empirical JER$^0$ for $B$ draws. If this is not the case for some $B$, then $\mathbf{t}^B_\alpha$ is set to the null family and the result holds.

By Theorem \ref{thm:PIdistrib} we have for all $\mathbf{t}$ that $\operatorname{JER}{\mathbf(\mathbf{t})} \leq \operatorname{JER}^0{\mathbf(\mathbf{t})}$. We can write:
\begin{align*}
  \operatorname{JER}^0{\mathbf(\mathbf{t})}
  & = \widehat{\operatorname{JER}}^0_B(\mathbf{t}) + \left( \operatorname{JER}^0{\mathbf(\mathbf{t})} - \widehat{\operatorname{JER}}^0_B(\mathbf{t}) \right) \\
  & = \widehat{\operatorname{JER}}^0_B(\mathbf{t}) + O_P(1/\sqrt{B})
\end{align*}
by Lemma~\ref{lem:JERpi1}.
Applying the above to $\mathbf{t} = \mathbf{t}^B_\alpha$ yields the desired result since $ \widehat{\operatorname{JER}}^0_{B}(\mathbf{t}^B_\alpha) \leq \alpha$ by definition.
\end{proof}

\newappendix{Algorithms}
Algorithm \ref{alg:nullp} describes the procedure to obtain samples from the joint distribution $(\pi^{0}_k)_k$. This is useful to compute the empirical JER of Equation \ref{eq:empiricalJER} via \ref{alg:estimateJER} and in turn to perform calibration which is described in Algorithm \ref{alg:calibration}. Once calibration is performed, inference can be performed using Algorithm \ref{alg:inference}. The FDP of resulting regions is provably controlled thanks to Theorem \ref{thm:JERpi2}.

\begin{algorithm}[H]
   \caption{\textbf{Sampling from the joint distribution of $\pi$ statistics under the null} according to Theorem \ref{thm:PIdistrib}.}
   \label{alg:nullp}
{\bfseries Input:} $B$ the number of MC draws; $p$ the number of variables\\
{\bfseries Output:} $\mathbf{\Pi}_0 \in [0, 1]^{B \times p}$ a matrix of $\pi^0$ statistics\\
$\mathbf{\Pi}_0 \gets$ zeros$(B, p)$\\
  \For{$b \in [1, B]$}{
      $\boldsymbol{\chi} \gets$ draw\_random\_vector$(\{-1, 1\}^{p})$ \Comment{Draw signs}\\
      $Z = 0$ \Comment{Initialize count}\\
      \For{$j \in [1, p]$}{
        \If{$\boldsymbol{\chi}[j] < 0$}{
          $\mathbf{\Pi}_0[b][j] \gets 1$\\
          $Z \gets Z + 1$ \Comment{Increment $Z$}\\
        }
        \Else{$\mathbf{\Pi}_{0}[b][j] \gets \frac{1 + Z}{p}$}
      }
      }
  $\mathbf{\Pi}_0 \gets$ sort\_lines($\mathbf{\Pi}_0$)
  \Comment{Sort samples}\\
{\bfseries Return} $\mathbf{\Pi}_0$
\end{algorithm}

\begin{algorithm}[H]
\caption{\textbf{Computing the Empirical JER.} The empirical JER is computed for a given threshold family and a matrix of $\pi^0$ statistics. This algorithm is similar to Algorithm 3 of \cite{blain2022notip}.}
\label{alg:estimateJER}
{\bfseries Input:} $\mathbf{\Pi}_0$ a matrix of $\pi^0$ statistics; $\mathbf{t}$ a threshold family; $k_{max}$ the size of the threshold family
{\bfseries Output:} $\widehat{\operatorname{JER}}$, the empirical JER of threshold family $\mathbf{t}$\\
 ($B$, p) $\gets$ shape($\mathbf{\Pi}_0$)\\
  $\widehat{\operatorname{JER}} \gets 0$\\
  \For{$b \in [1, B]$}{
    \For{$i \in [1, k_{max}]$}{
     diff[i] $\gets \mathbf{\Pi}_0[b'][i] -\mathbf{t}[i]$
    \\ \Comment{Check JER control at rank $i$}
    }
  \If{$\mathrm{min(diff)} < 0$}{
   $\widehat{\operatorname{JER}} \gets \widehat{\operatorname{JER}} + \frac{1}{B}$
  \\ \Comment{Increment risk if JER control event is violated}
  }
  }
 {\bfseries Return} $\widehat{\operatorname{JER}}$
\end{algorithm}

\begin{algorithm}[H]
   \caption{\textbf{Performing calibration on $\pi$-statistics}. First, we use Theorem \ref{thm:PIdistrib} to build a suitable template and estimate the JER of each candidate threshold family. Then, we perform calibration to select the least conservative possible threshold family that controls the JER at a given level $\alpha$.}
   \label{alg:calibration}
 {\bfseries Input:} $\alpha$ the desired FDP coverage; $B$ the number of MC draws for JER estimation; $B'$ the number of candidate threshold families\\
 {\bfseries Output:} $\mathbf{t}_{\alpha}$ the calibrated threshold family at level $\alpha$\\
   $\mathbf{\Pi}_0 \gets$ draw\_null\_$\pi(B, p)$ \Comment{Algorithm \ref{alg:nullp}}\\
   $\mathbf{\Pi}_0^{'} \gets$ draw\_null\_$\pi(B', p)$\\
  \For{$b' \in [1, B']$}{
   $\mathbf{T}[b'] \gets$ quantiles($\mathbf{\Pi}_0^{'}$, $\frac{b'}{B'}$) \Comment{Build template}\\
   $\widehat{\operatorname{JER}}_{b'} \gets \text{empirical\_jer}(\mathbf{\Pi}_0,  \mathbf{T}[b'])$ \Comment{Apply Algorithm \ref{alg:estimateJER} for each family}\\
  }
     $b'_{cal} \gets$ $\operatorname{max}\{b' \in [1, B']$ s.t. $\widehat{\operatorname{JER}}_{b'} \leq \alpha\}$ \Comment{Perform calibration}\\
  
   $\mathbf{t}_{\alpha} \gets \mathbf{T}[b'_{cal}]$\\

 {\bfseries Return} $\mathbf{t}_{\alpha}$
\end{algorithm}

\begin{algorithm}[h]
   \caption{\textbf{Performing inference via Knockoffs and calibration.} We compute the largest possible region that satisfies the required FDP level $q$ using the JER controlling family computed via Algorithm \ref{alg:calibration}. The bound $V^{\mathbf{t}_{\alpha}}$ is computed from $\mathbf{\pi}$ using Equation \ref{eq:bound}.}
   \label{alg:inference}

 {\bfseries Input:} $\mathbf{X}$ the input data; $\mathbf{y}$ the target variable; $q$ the maximum tolerable FDP; $\mathbf{t}_{\alpha}$ the calibrated threshold family at level $\alpha$\\
 {\bfseries Output:} $\hat{S}$ the selected variables\\
   $n, p \gets$ shape($\mathbf{X}$) \Comment{n samples, p variables}\\
   $\mathbf{\tilde{X}} \gets \text{sample\_Knockoffs}(\mathbf{X})$\\
   $\mathbf{W} \gets LCD(\mathbf{X}, \mathbf{\tilde{X}}, \mathbf{y})$ \Comment{Compute $\mathbf{W}$}\\
   $\boldsymbol{\pi} \gets \text{compute\_proportion}(\mathbf{W})$ \Comment{Equation \eqref{eq:proportion}}\\
   $\hat{S} \gets \displaystyle \max_{S} \{|S| \quad s.t. \quad \frac{V^{\mathbf{t}_{\alpha}}(S)}{|S|} \leq q\}$ \Comment{Find largest admissible region}\\
   \Comment{$V^{\mathbf{t}_{\alpha}}(S)$ depends on $\boldsymbol{\pi}$ }

 {\bfseries Return} $\hat{S}$
\end{algorithm}

\newappendix{Additional simulation results}

\newsubappendix{A harder inference setup}

\begin{figure}[h!]
\centering
\includegraphics[width=0.7\linewidth]{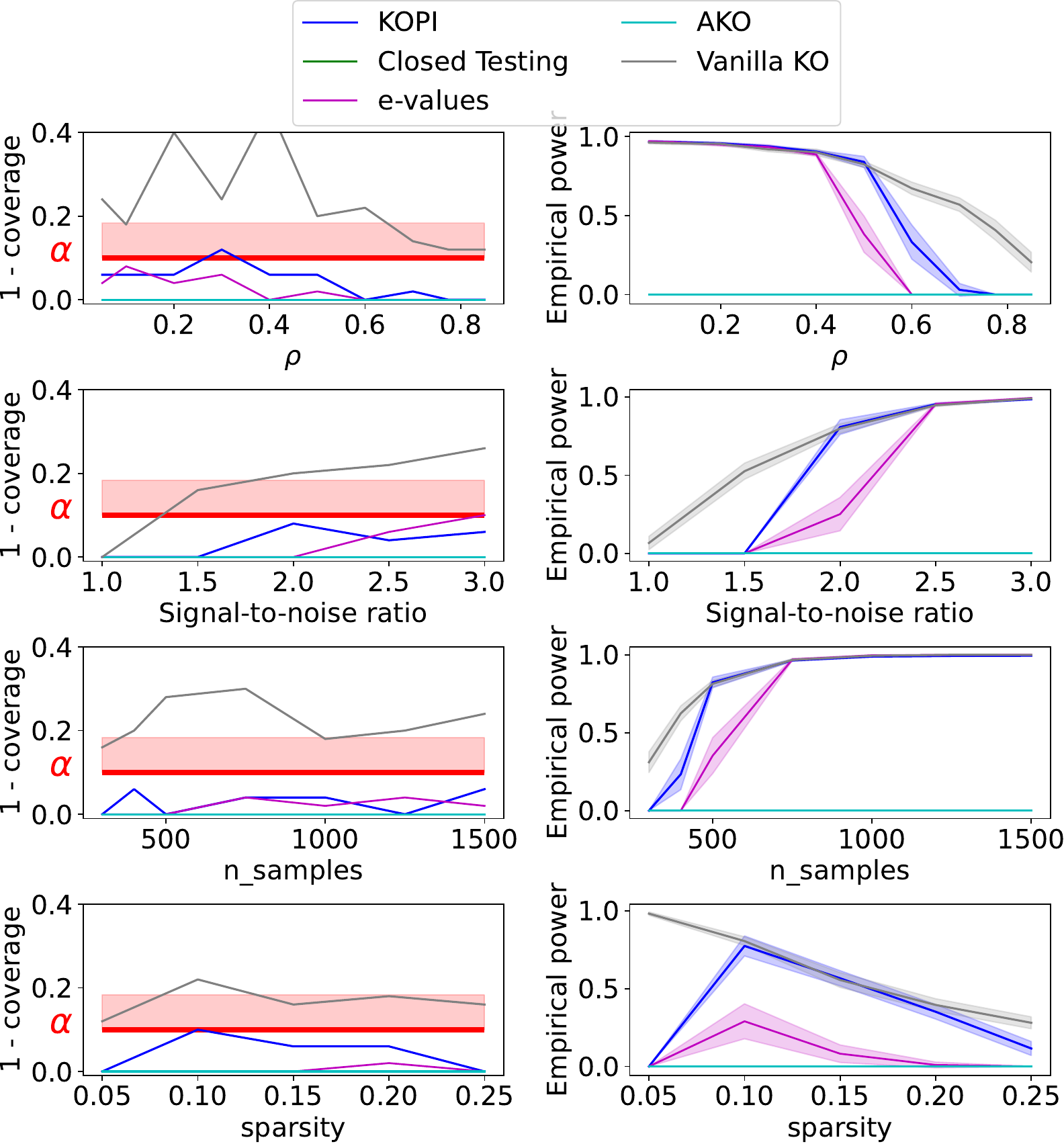}
\caption{\textbf{FDP bound coverage at level $\alpha$ and empirical Power for 50 simulation runs and five different methods.} The five methods are Vanilla Knockoffs, aggregated Knockoffs using e-values, aggregated Knockoffs using quantile-aggregation, KOPI and Knockoff inference via Closed Testing. We use 50 Knockoffs draws and the following simulation setting $\alpha = 0.1, q = 0.05, p = 500$. Each row represents a varying parameter with the left panel displaying FDP coverage and the right panel displaying power. The red line and associated error bands represent the acceptable limits for FDP bound coverage. Notice that KOPI consistently outperforms all other methods while retaining FDP control.\looseness=-1}
\label{fig:sim2}
\end{figure}

We evaluated the performance of all five methods in the more challenging setting $q = 0.05$ instead of using $q = 0.1$. 
The results are presented in Fig. \ref{fig:sim2}.
In this setting, AKO and Closed Testing are always powerless and aggregation via e-values suffers from a lack of power in most cases. Vanilla Knockoffs exhibit satisfactory power but consistently fail to control the FDP. KOPI preserves FDP control and yields acceptable power.

\newsubappendix{Impact of aggregation scheme choice}

While the theoretical guarantees we obtain hold for all choices of aggregation schemes, these hyperparameter impacts the power of KOPI. To assess this, we use the same simulated data setup as in Figure \ref{fig:sim1} to compare four aggregation schemes: arithmetic mean, geometric mean, harmonic mean and quantile aggregation.

Importantly, we first check that the FDP is controlled for all types of aggregation and in all settings considered by reporting the bound non-coverage. We use three settings of varying difficulty, parametrized by the correlation level $\rho$ and use $\alpha = 0.1, q = 0.1$:

\begin{table}[h!]
    \centering
    \begin{tabular}{l|l|l|l|l}
        ~ & Harmonic & Arithmetic & Geometric & Quantile aggregation \\ \hline
          $\rho = 0.5$ &  $10 \%$  &  $0 \%$  &  $2 \%$  &  $10 \%$  \\ \hline
          $\rho = 0.6$  &  $2 \%$  &  $0 \%$  &  $0 \%$  &  $4 \%$  \\ \hline
          $\rho = 0.7$  &  $2 \%$  &  $0 \%$  &  $0 \%$  &  $0 \%$  \\
    \end{tabular}
    \caption{\textbf{FDP control of KOPI for four aggregation schemes and three different correlation levels.} Note that FDP control is maintained in all scenarios which is coherent with the result obtained in Theorem \ref{thm:JERpi2}.}
    \label{tab:AggSimuFDP}
\end{table}

The FDP is indeed controlled in all cases since non-coverage never exceeds the chosen level $\alpha = 10\%$ as seen in Table \ref{tab:AggSimuFDP}. This is coherent with the theoretical guarantees we obtain in Theorem \ref{thm:JERpi2}. We now report the average power to benchmark aggregation schemes:

\begin{table}[h!]
    \centering
    \begin{tabular}{l|l|l|l|l}
        ~ & Harmonic & Arithmetic & Geometric & Quantile aggregation \\ \hline
          $\rho = 0.5$ &  $\mathbf{0.91}$  &  $0.77$  &  $0.87$  &  $0.90$  \\ \hline
          $\rho = 0.6$  &  $\mathbf{0.83}$  &  $0.58$  &  $0.77$  &  $\mathbf{0.83}$  \\ \hline
          $\rho = 0.7$  &  $\mathbf{0.72}$  &  $0.39$  &  $0.61$  &  $\mathbf{0.72}$  \\
    \end{tabular}
    \caption{\textbf{Empirical power of KOPI for four aggregation schemes and three different correlation levels.} Note that harmonic mean aggregation consistently outperforms arithmetic aggregation and geometric aggregation. Quantile aggregation performs similarly to harmonic aggregation.}
    \label{tab:AggSimuPOW}
\end{table}

Note that harmonic mean aggregation outperforms arithmetic and geometric mean consistently and performs similarly to quantile aggregation as seen in Table \ref{tab:AggSimuPOW}.

\newappendix{Details and results on genomic data}
\label{sec:GenomicsExpe}

\newsubappendix{Lymphomatic leukemia mutation classification}
Differential gene expression studies aim at identifying genes whose activity differs significantly between two (or more) populations, based on a sample of measurements from individuals from these populations.
The activity of a gene is usually quantified by its level of expression in the cell.
We consider a microarray data set studied in \cite{bourgon2010independent} that consists of expression measurements for biological samples from $n = 79$ individuals with B-cell acute lymphoblastic leukemia (ALL):
$37$ of these individuals harbor a specific mutation called  BCR/ABL, while the remaining $42$ do not.
Our goal here is to identify, from this sample, genes for which there is a difference in the mean expression level between the mutated and non-mutated populations.
We focus on the $p=90$ genes on chromosome 7 whose individual standard deviation is above $0.5$.

The genes selected by different Knockoffs-based methods are summarized in Table \ref{tab:GenomicsExpe}. Stability selection criteria analogous to \cite{luo2022improving, ren2021derandomizing} are displayed. Note that the selection made by KOPI is more robust than that of Vanilla Knockoffs: $4$ genes are selected in nearly all runs by KOPI, while none are selected as frequently by Vanilla Knockoffs.
Conversely, KOPI only selects $2$ genes in less than $50\%$ of all runs compared to $18$ for Vanilla Knockoffs. 
This confirms that error control guarantees of KOPI, together with the stability brought by aggregation, lead to avoiding most spurious/non-reproducible detections. Besides KOPI and Vanilla Knockoffs, all other methods are powerless in all runs.
\begin{table}[h!]
\begin{tabular}{l|l|l|l|l|l}
                                                                    & KOPI                     & Vanilla KO                & e-values & Closed Testing & AKO \\ \hline
\multicolumn{1}{l|}{Selected in \textgreater 90\% of runs}         & {\color[HTML]{32CB00} 4} & {\color[HTML]{FE0000} 0}  & 0        & 0              & 0   \\ \hline
\multicolumn{1}{l|}{Selected in \textgreater 50\% of runs}         & 6                        & 6                         & 0        & 0              & 0   \\ \hline
\multicolumn{1}{l|}{Spurious detections (\textless{}50\% of runs)} & {\color[HTML]{32CB00} 2} & {\color[HTML]{FE0000} 18} & 0        & 0              & 0   \\
\end{tabular}
\caption{\textbf{Stability selection criteria for 5 Knockoffs-based methods on "Lymphomatic leukemia mutation" genomic data.} Note that KOPI displays a very stable selection set across all runs with $4$ genes present in $>90 \%$ of runs. KOPI also avoids most spurious discoveries, as only $2$ genes are selected less than $50\%$ of the time, compared to $18$ genes using Vanilla Knockoffs. The $6$ genes selected more than $50\%$ of the time by KOPI and Vanilla Knockoffs are the same. All other Knockoffs-based methods are powerless in all runs.}
\label{tab:GenomicsExpe}
\end{table}

\newsubappendix{Colon vs Kidney classification}

We also considered an additional genomic dataset to reproduce these results with a larger number of samples.  The dataset we used is part of \textbf{GEMLeR (Gene Expression Machine Learning Repository)} \cite{stiglic2010stability}, a collection of gene expression datasets that can be used to benchmark ML methods on genomics data.

We chose the "Colon vs Kidney" dataset: this is a binary classification dataset where the goal is to distinguish cancerous tissue from two different organs (Colon and Kidney) using gene expression data. This dataset comprises $546$ samples and $10936$ genes. To make the problem tractable for Knockoffs-based methods we perform dimensionality reduction to select the $546$ genes that have the largest variance. Then, \textbf{we run all Knockoffs-based methods 50 times} and report the selected genes.

\begin{table}[h!]
\begin{tabular}{l|l|l|l|l|l}
                                                                    & KOPI                     & Vanilla KO                & e-values & Closed Testing & AKO \\ \hline
\multicolumn{1}{l|}{Selected in \textgreater 90\% of runs}         & {\color[HTML]{32CB00} 21} & {\color[HTML]{FE0000} 0}  & 0        & 0              & 0   \\ \hline
\multicolumn{1}{l|}{Selected in \textgreater 50\% of runs}         & 22                        & 25                         & 0        & 0              & 0   \\ \hline
\multicolumn{1}{l|}{Spurious detections (\textless{}50\% of runs)} & {\color[HTML]{32CB00} 7} & {\color[HTML]{FE0000} 34} & {\color[HTML]{FE0000} 20}        & 0              & 0   \\
\end{tabular}
\caption{\textbf{Stability selection criteria for 5 Knockoffs-based methods on "Colon vs Kidney" genomic data.} Note that KOPI displays a very stable selection set across all runs with $21$ genes present in $>90 \%$ of runs. KOPI also avoids most spurious discoveries, as only $7$ genes are selected less than $50\%$ of the time, compared to $34$ genes using Vanilla Knockoffs and $20$ using e-values. All other Knockoffs-based methods are powerless in all runs.}
\label{tab:GenomicsExpev2}
\end{table}

The genes selected by different Knockoffs-based methods are summarized in Table \ref{tab:GenomicsExpev2}. Stability selection criteria analogous to \cite{luo2022improving, ren2021derandomizing} are displayed. Note that the selection made by KOPI is more robust than that of Vanilla Knockoffs: $21$ genes are selected in nearly all runs by KOPI, while none are selected as frequently by Vanilla Knockoffs.
Conversely, KOPI only selects $7$ genes in less than $50\%$ of all runs compared to $34$ for Vanilla Knockoffs and $20$ for e-values aggregation.

\newappendix{Details and results on HCP data}
\label{sec:HCPappendix}

\newsubappendix{HCP dataset}

We use the HCP900 task-evoked fMRI dataset \citep{van2012}, in
which we take the masked $2$\,mm resolution z-statistics maps of the $778$ subjects from $7$
tasks to solve binary regression problems, namely predicting which condition is associated with the brain image: emotion (\emph{emotional face} vs \emph{shape outline}), gambling (\emph{reward} vs \emph{loss}), language (\emph{story} vs
\emph{math}), motor hand (\emph{left} vs \emph{right} hand), motor foot (\emph{left} vs \emph{right} foot), relational (\emph{relational} vs \emph{match}) and social (\emph{mental interaction} vs \emph{random interaction}).

We consider the fixed-effect maps (average across right-left and left-right phase encoding schemes) for each condition, yielding one
image per subject per condition (which corresponds to two images per
subject for each classification problem).
Then, for each problem, the number of samples available is $1556$
($=2 \times 778$) and the number of voxels is $156~374$ after
gray-matter masking.
Dimension reduction was carried out using Ward parcellation scheme to $1k$ clusters, which is known to yield spatially homogeneous regions \cite{thirion:ward}.
The signal is then averaged per cluster, yielding a reduced design matrix $\mathbf{X}$ for the problem.

\newsubappendix{Brain data are non-Gaussian}

In the synthetic data experiments we used the Gaussian Knockoff generation process described in \citealp{candes2018panning}. However, fMRI brain maps can be heavily non-Gaussian. In turn, Gaussian Knockoffs cannot satisfy the Knockoffs exchangeability assumption and any statistical control on False Discoveries is rendered spurious.

\begin{figure}[h!]
\centering
\includegraphics[width=0.9\linewidth]{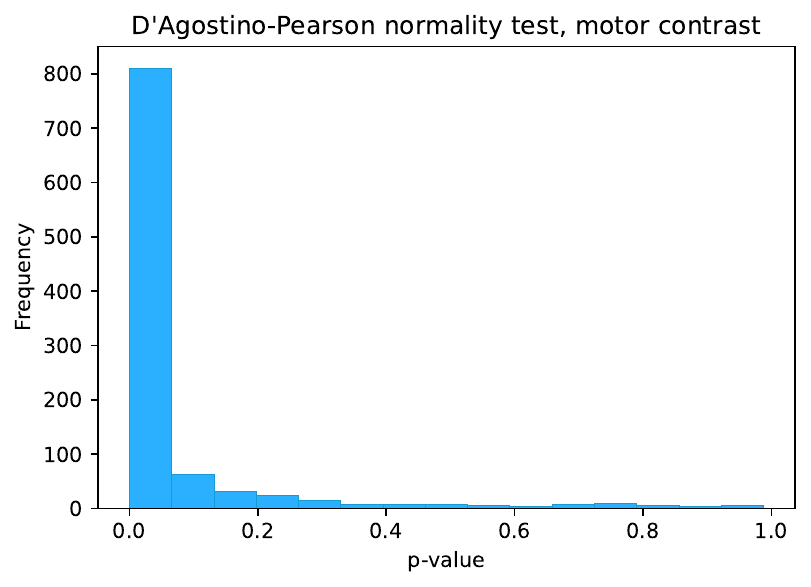}
\caption{\textbf{D'Agostino-Pearson normality test for a motor contrast.} We perform a normality test for each cluster amongst the $1000$ present for the motor foot contrast. The distribution of the normality test $p$-values indicates strong non-normality in fMRI data.}
\label{fig:nongaussian}
\end{figure}

To build non-Gaussian Knockoffs, we use a linear variant of the Sequential Conditional Independent Pairs (SCIP) algorithm of \citealp{candes2018panning}:

\begin{algorithm}[h!]
\caption{Generating Non-Gaussian Knockoffs using the Sequential Conditional Independent Pairs algorithm of \citealp{candes2018panning}.}
\For{$j \in [1, p]$}{Fit a Lasso model on $(\mathbf{X_{-j}}, X_j)$\\
Compute the residual $\epsilon_j = X_j - \mathbf{X_{-j}}\boldsymbol{\widehat{\beta}}_j$}

\For{$j \in [1, p]$}{$\text{Sample } \tilde{X}_j \text { from } \mathbf{X_{-j}}\boldsymbol{\widehat{\beta}}_j + \epsilon_{\rho{(j)}}$ \Comment{$\rho$ is a random ordering of $[1, p]$}} 

{\bfseries Return} $\mathbf{\tilde{X}_{1: p}}$
\end{algorithm}

\newsubappendix{Additional results}

\begin{figure}[h]
\centering
\includegraphics[width=\linewidth]{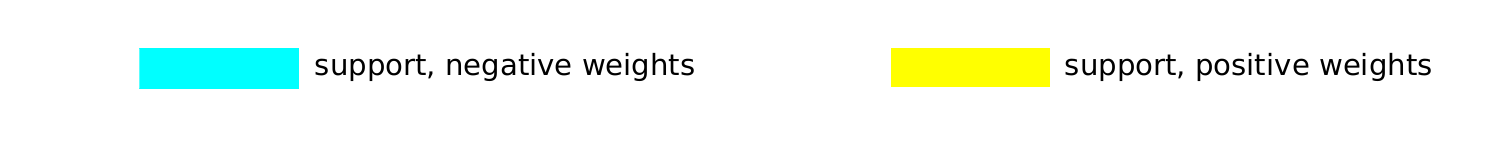}
\includegraphics[width=0.8\linewidth]{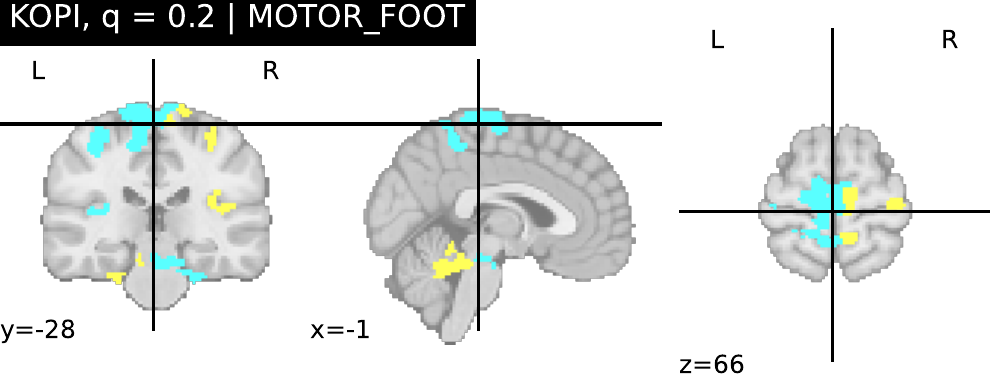}
\includegraphics[width=0.8\linewidth]{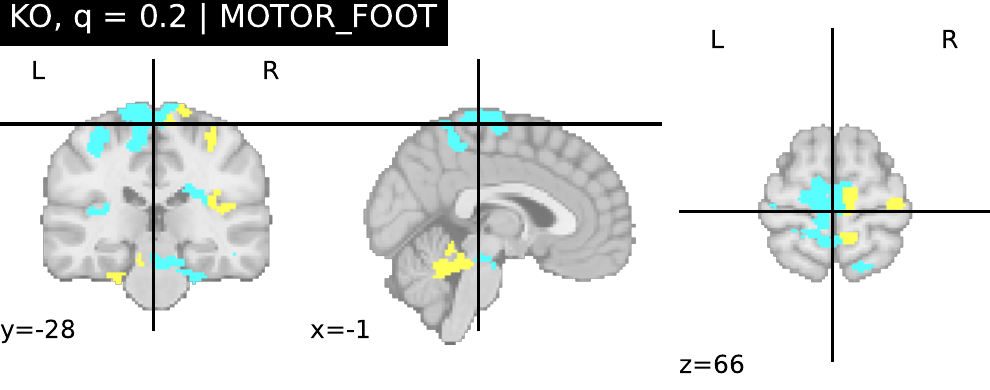}
\includegraphics[width=0.8\linewidth]{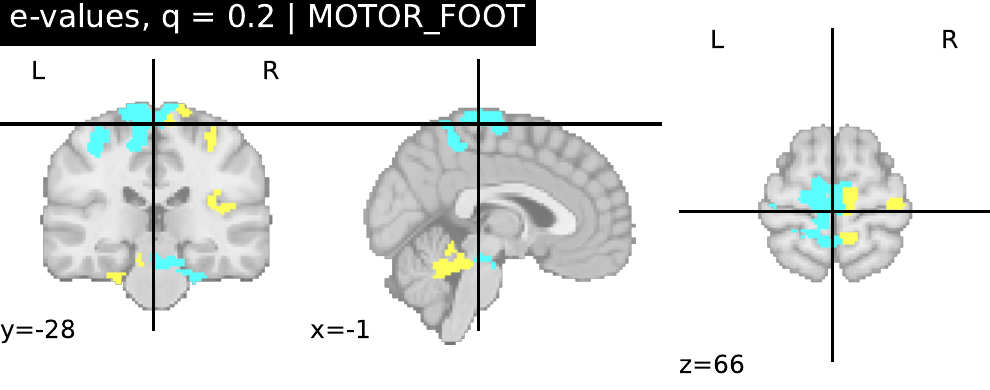}
\includegraphics[width=0.8\linewidth]{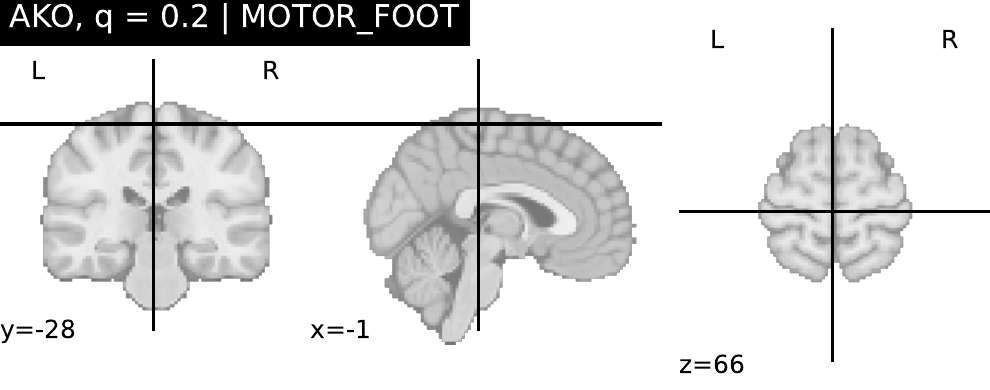}
\caption{\textbf{Brain mapping on motor contrast using Knockoffs-based methods.} Among the five methods considered in this paper --Vanilla Knockoffs, aggregated Knockoffs using e-values, aggregated Knockoffs using quantile-aggregation (AKO), KOPI and Knockoff inference via Closed Testing-- only Vanilla Knockoffs, e-values and KOPI yield discoveries, plotted above. All other methods are powerless. We use 50 Knockoffs draws and $\alpha = 0.1$ and $q = 0.2$. Each figure represents the region returned by a given method. Vanilla Knockoffs yield 17 regions, KOPI: 24 regions and e-values: 18 regions.}
\label{fig:fMRI1}

\end{figure}

\begin{figure}[h]
\centering
\includegraphics[width=\linewidth]{figures/rect.pdf}
\includegraphics[width=0.8\linewidth]{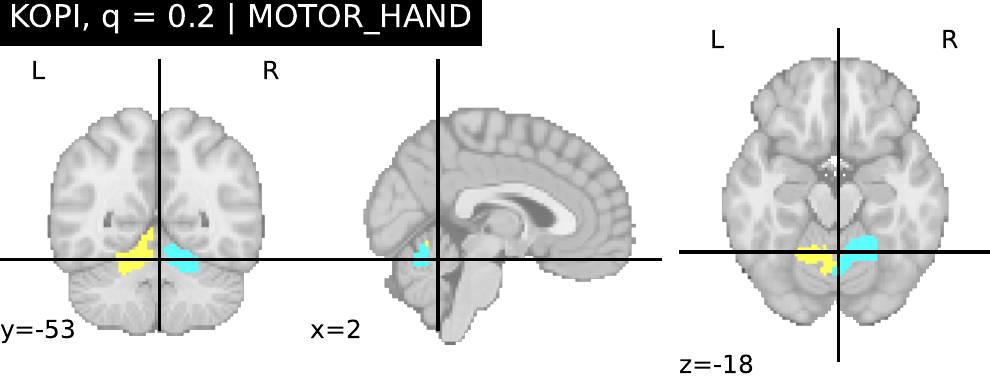}
\includegraphics[width=0.8\linewidth]{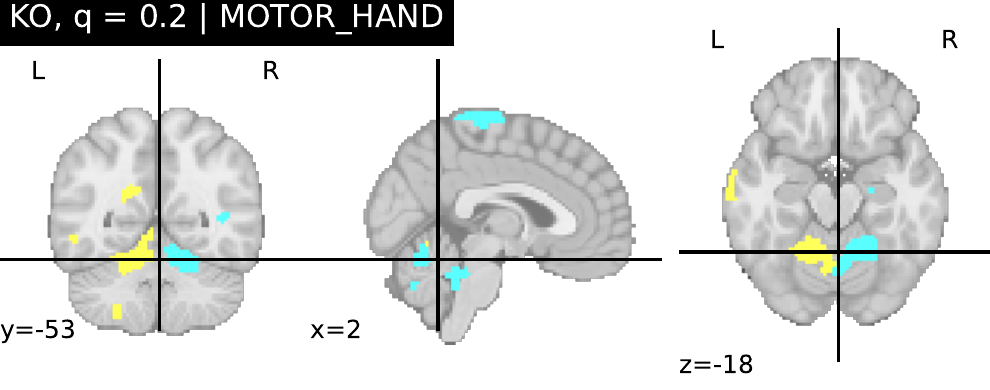}
\includegraphics[width=0.8\linewidth]{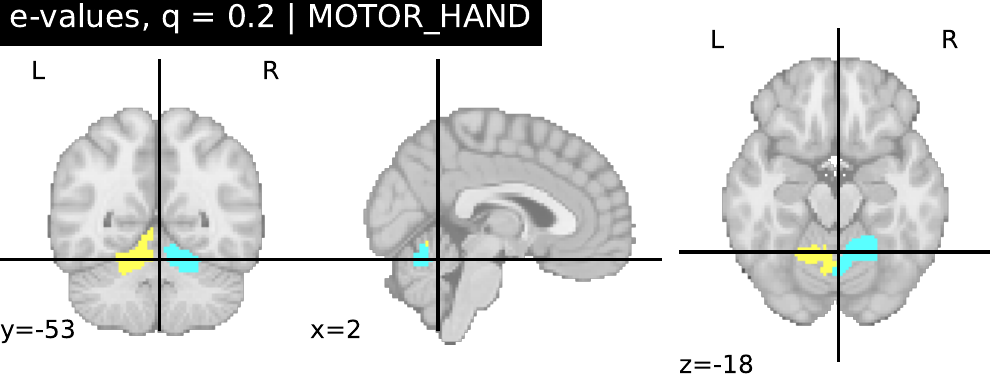}
\includegraphics[width=0.8\linewidth]{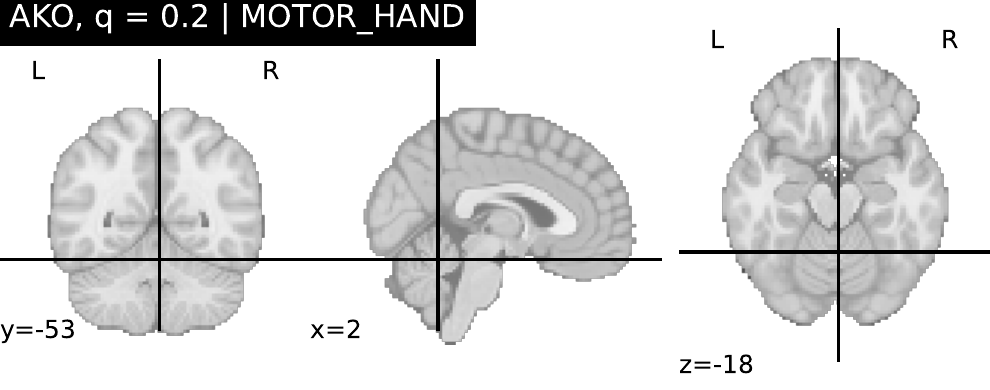}
\caption{\textbf{Brain mapping on motor hand contrast using Knockoffs-based methods.} Among the five methods considered in this paper --Vanilla Knockoffs, aggregated Knockoffs using e-values, aggregated Knockoffs using quantile-aggregation (AKO), KOPI and Knockoff inference via Closed Testing-- only Vanilla Knockoffs, e-values and KOPI yield discoveries, plotted above. All other methods are powerless. We use 50 Knockoffs draws and $\alpha = 0.1$ and $q = 0.2$. Each figure represents the region returned by a given method. Vanilla Knockoffs yield 11 regions, KOPI, 10 regions and e-values 11 regions.}
\label{fig:fMRI2}

\end{figure}

\begin{figure}[h]
\centering
\includegraphics[width=\linewidth]{figures/rect.pdf}
\includegraphics[width=0.8\linewidth]{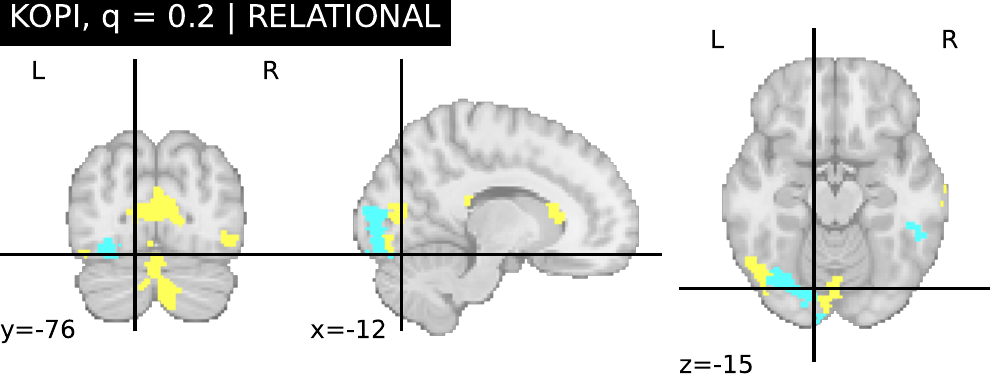}
\includegraphics[width=0.8\linewidth]{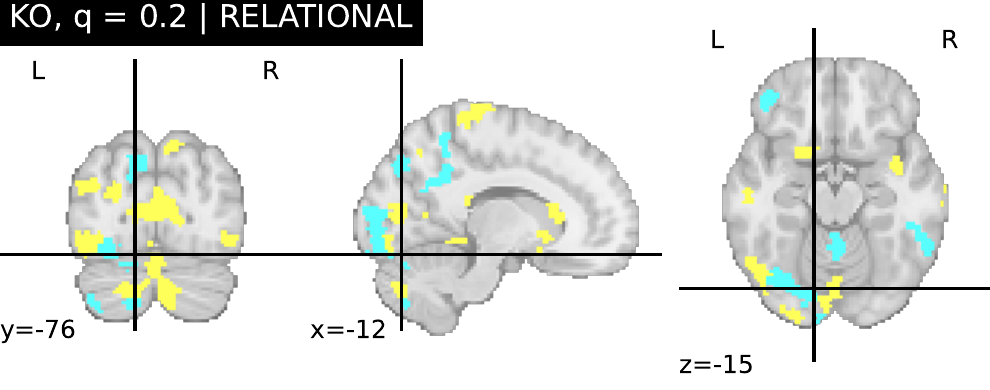}
\includegraphics[width=0.8\linewidth]{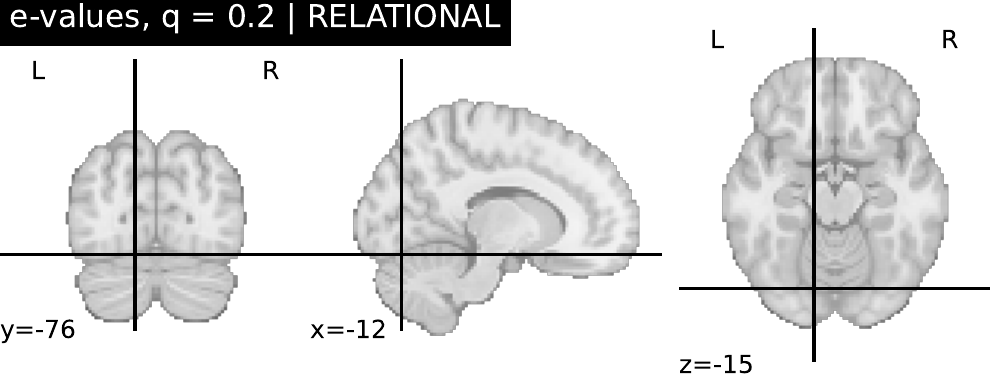}
\includegraphics[width=0.8\linewidth]{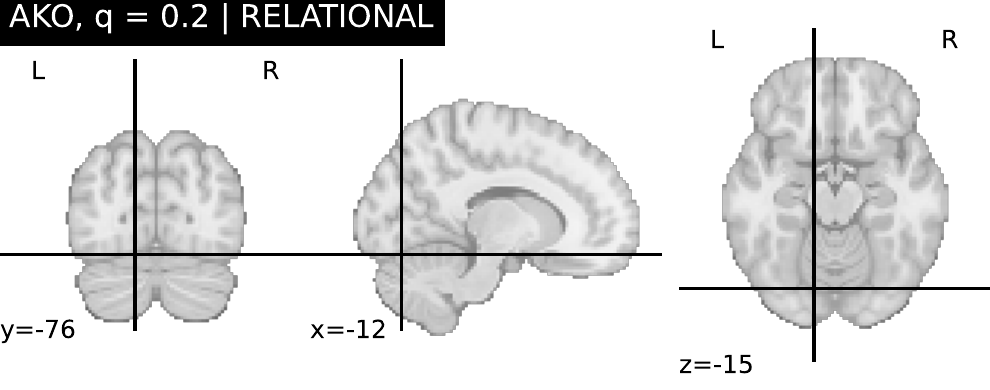}
\caption{\textbf{Brain mapping on the HCP Relational task using Knockoffs-based methods.} Among the five methods considered in this paper --Vanilla Knockoffs, aggregated Knockoffs using e-values, aggregated Knockoffs using quantile-aggregation (AKO), KOPI and Knockoff inference via Closed Testing-- only Vanilla Knockoffs and KOPI yield discoveries, plotted above. All other methods are powerless. We use 50 Knockoffs draws, $\alpha = 0.1$ and $q = 0.2$. Each figure represents the region returned by a given method. Vanilla Knockoffs yield 58 regions and KOPI, 24 regions.}
\label{fig:fMRI3}

\end{figure}

\begin{figure}[h]
\centering
\includegraphics[width=\linewidth]{figures/rect.pdf}
\includegraphics[width=0.8\linewidth]{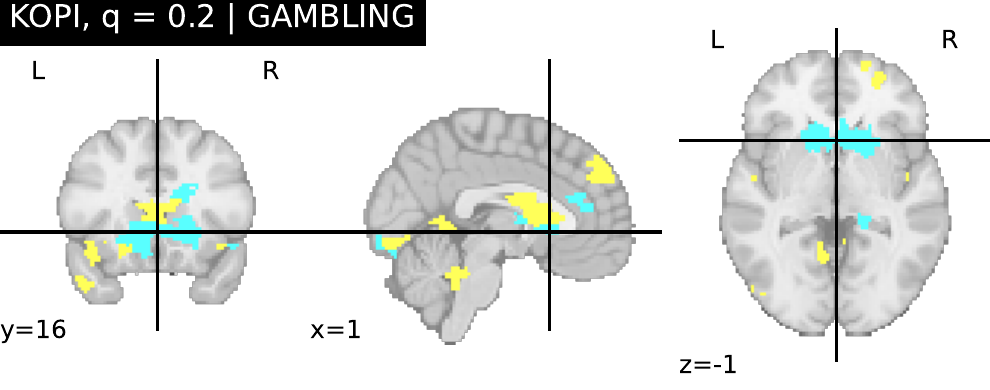}
\includegraphics[width=0.8\linewidth]{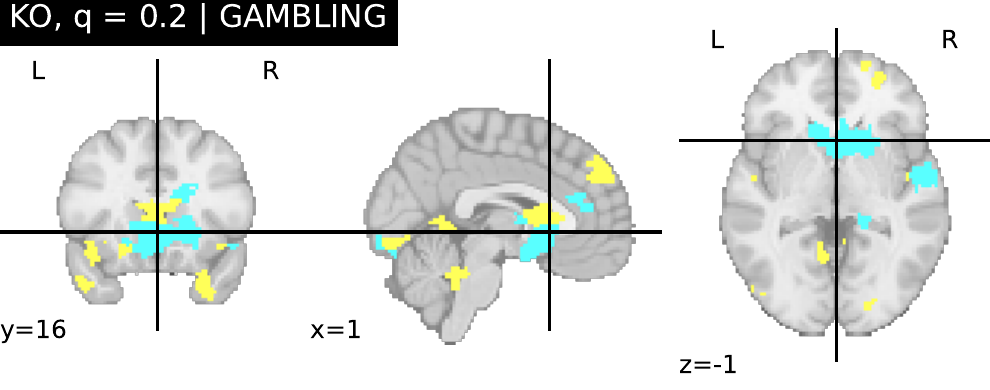}
\includegraphics[width=0.8\linewidth]{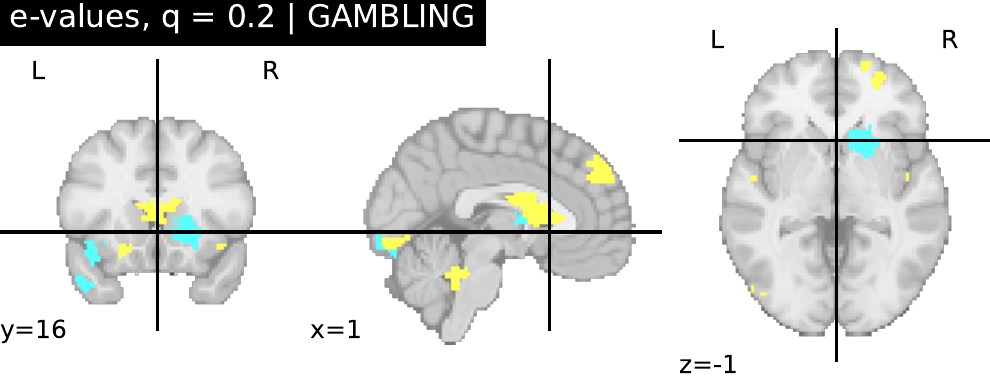}
\includegraphics[width=0.8\linewidth]{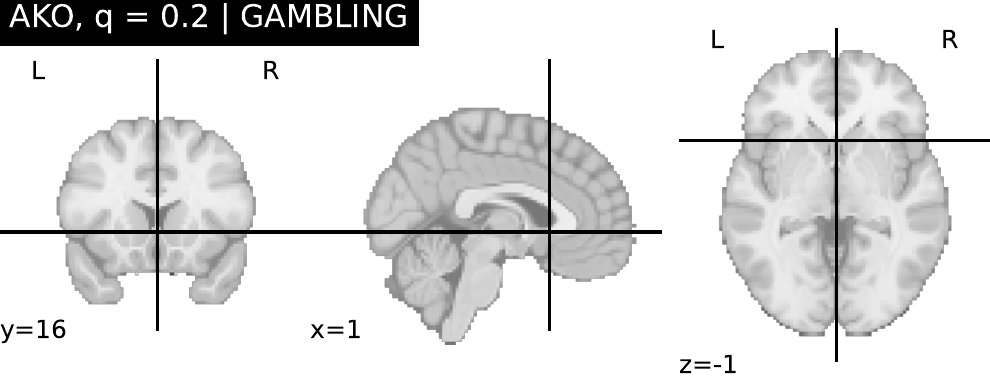}
\caption{\textbf{Brain mapping on HCP gambling task using Knockoffs-based methods.} Among the five methods considered in this paper --Vanilla Knockoffs, aggregated Knockoffs using e-values, aggregated Knockoffs using quantile-aggregation (AKO), KOPI and Knockoff inference via Closed Testing-- only Vanilla Knockoffs, KOPI and e-values aggregation yield discoveries, plotted above. All other methods are powerless. We use 50 Knockoffs draws, $\alpha = 0.1$ and $q = 0.2$. Each figure represents the region returned by a given method. Vanilla Knockoffs yield 57 regions, KOPI 57 regions, e-values aggregation, 19 regions.}
\label{fig:fMRI4}

\end{figure}

\begin{figure}[h]
\centering
\includegraphics[width=\linewidth]{figures/rect.pdf}
\includegraphics[width=0.8\linewidth]{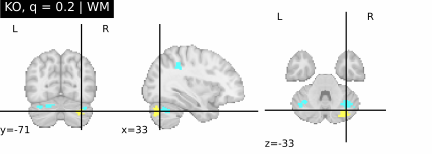}
\includegraphics[width=0.8\linewidth]{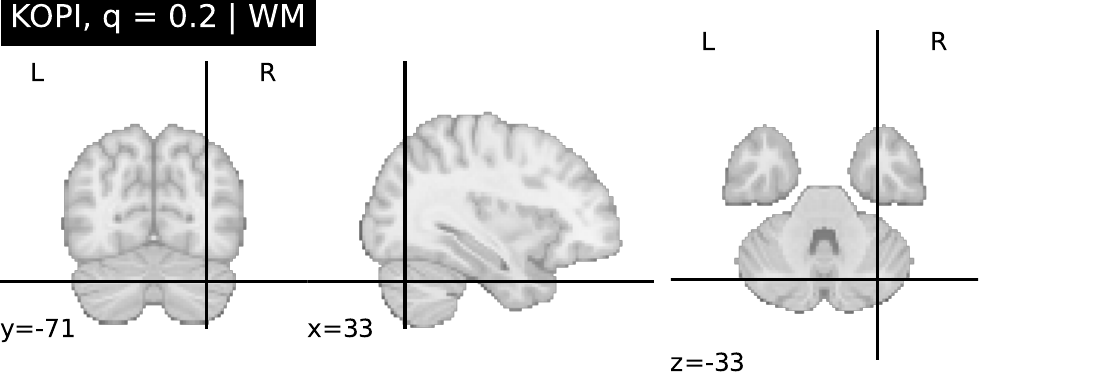}
\includegraphics[width=0.8\linewidth]{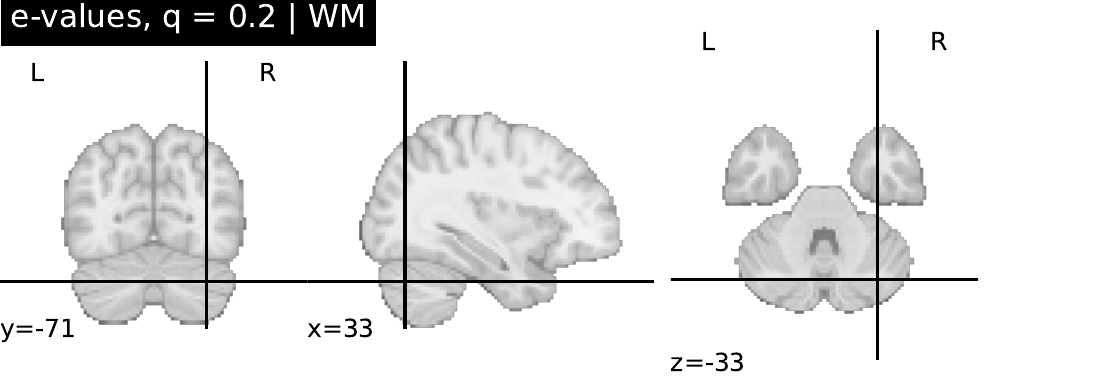}
\includegraphics[width=0.8\linewidth]{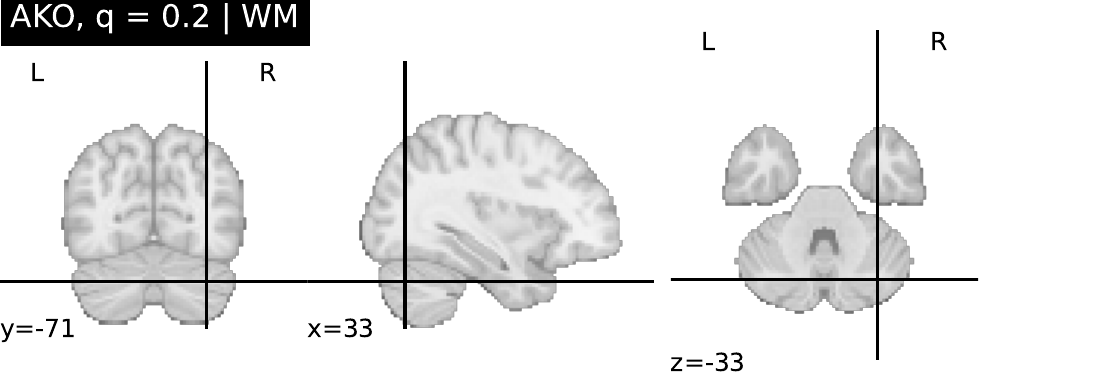}
\caption{\textbf{Brain mapping on HCP working memory task using Knockoffs-based methods.} Among the five methods considered in this paper --Vanilla Knockoffs, aggregated Knockoffs using e-values, aggregated Knockoffs using quantile-aggregation (AKO), KOPI and Knockoff inference via Closed Testing-- only Vanilla Knockoffs yields discoveries, plotted above. All other methods are powerless. We use 50 Knockoffs draws, $\alpha = 0.1$ and $q = 0.2$. Each figure represents the region returned by a given method. Vanilla Knockoffs yield 8 regions.}
\label{fig:fMRI5}

\end{figure}

\begin{figure}[h]
\centering
\includegraphics[width=\linewidth]{figures/rect.pdf}
\includegraphics[width=0.8\linewidth]{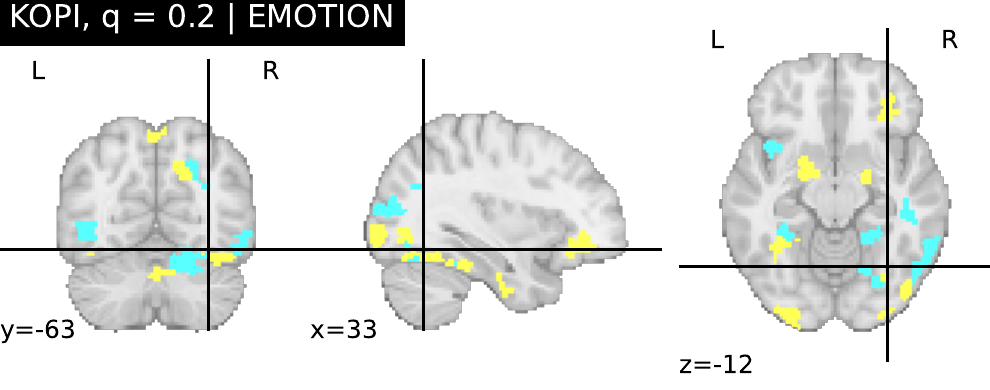}
\includegraphics[width=0.8\linewidth]{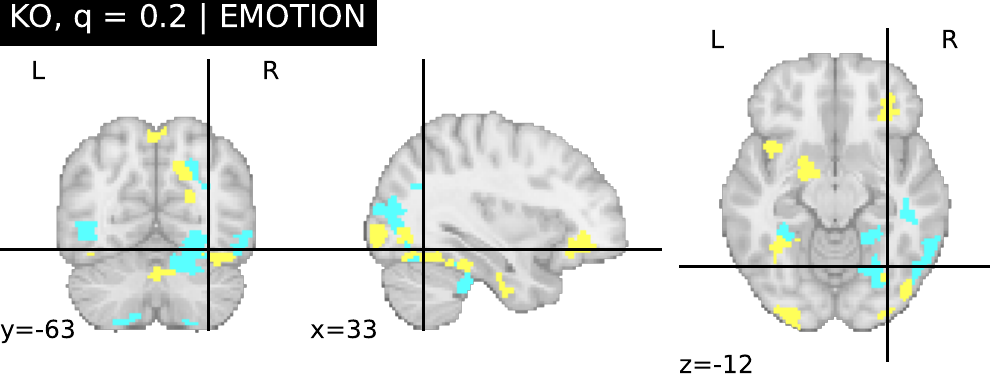}
\includegraphics[width=0.8\linewidth]{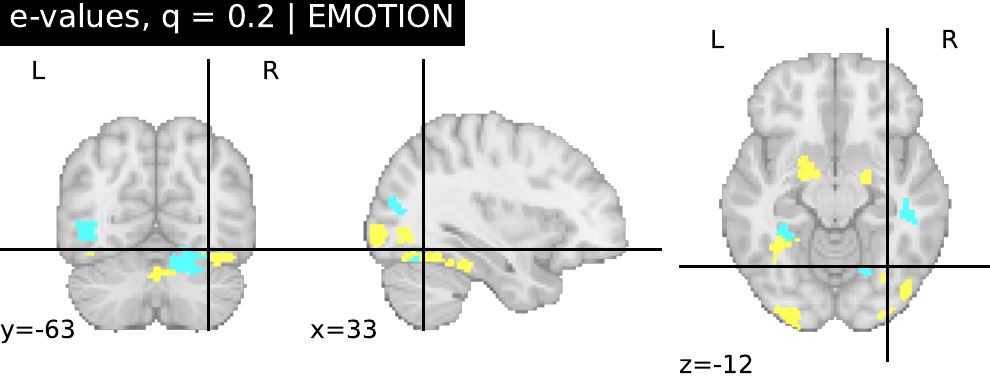}
\includegraphics[width=0.8\linewidth]{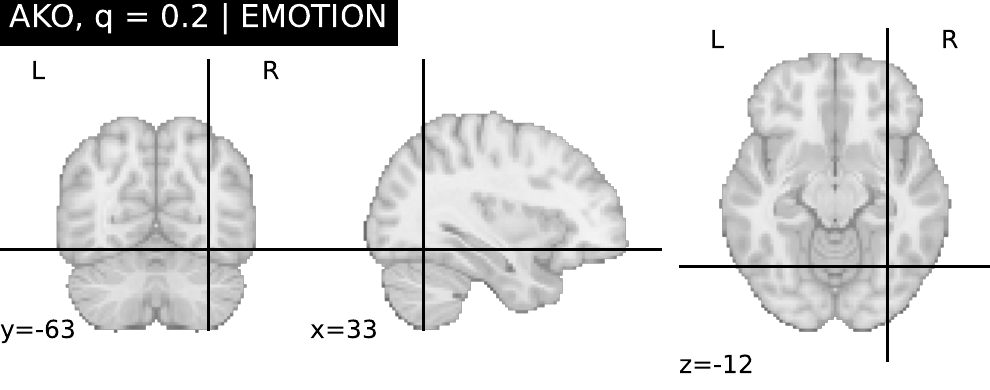}
\caption{\textbf{Brain mapping on HCP emotional task using Knockoffs-based methods.} Among the five methods considered in this paper --Vanilla Knockoffs, aggregated Knockoffs using e-values, aggregated Knockoffs using quantile-aggregation (AKO), KOPI and Knockoff inference via Closed Testing-- only Vanilla Knockoffs, KOPI and e-values aggregation yield discoveries, plotted above. All other methods are powerless. We use 50 Knockoffs draws, $\alpha = 0.1$ and $q = 0.2$. Each figure represents the region returned by a given method. Vanilla Knockoffs yield 22 regions, KOPI: 37 regions,  e-values aggregation: 20 regions.}
\label{fig:fMRI6}

\end{figure}

\begin{figure}[h]
\centering
\includegraphics[width=\linewidth]{figures/rect.pdf}
\includegraphics[width=0.8\linewidth]{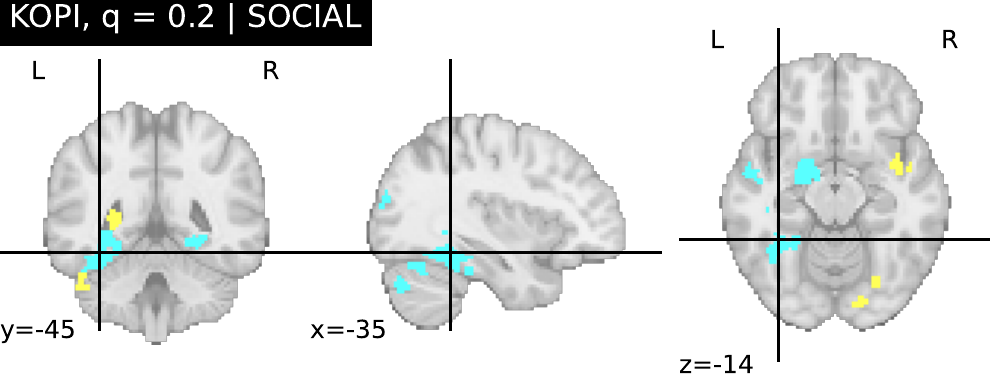}
\includegraphics[width=0.8\linewidth]{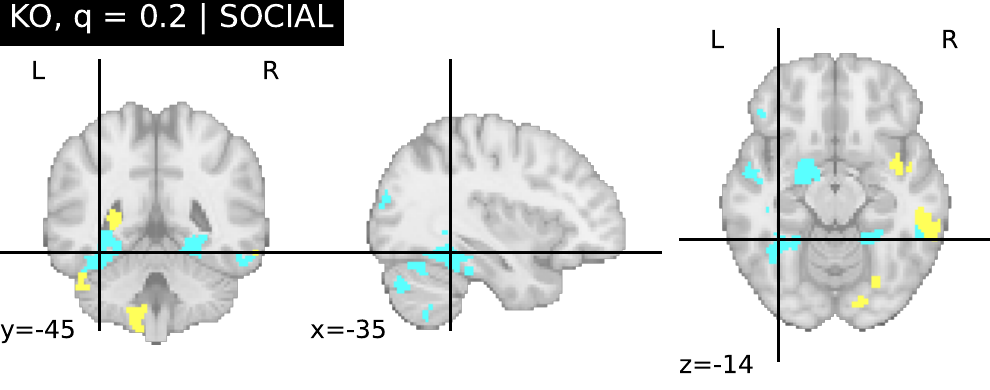}
\includegraphics[width=0.8\linewidth]{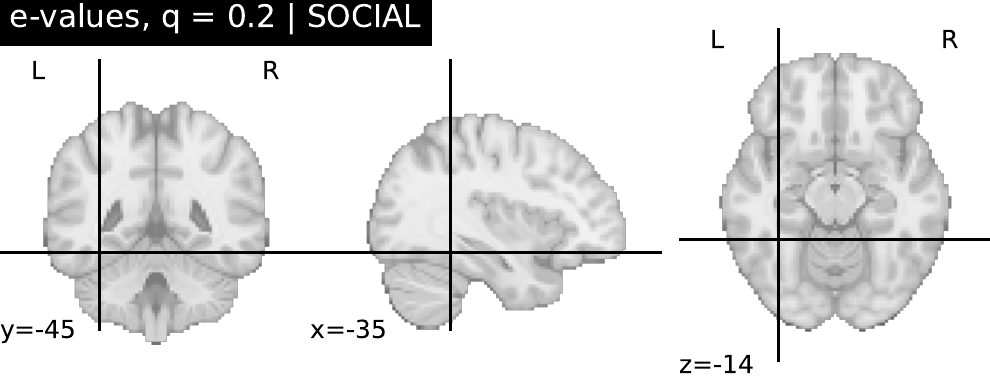}
\includegraphics[width=0.8\linewidth]{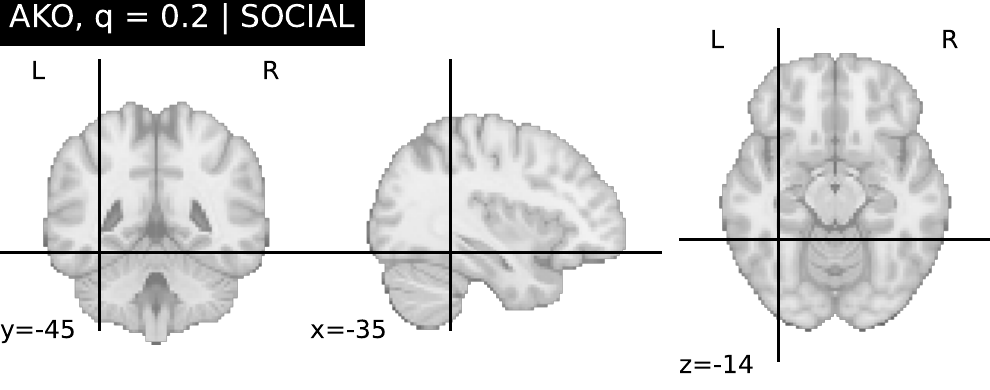}
\caption{\textbf{Brain mapping on HCP social task using Knockoffs-based methods.} Among the five methods considered in this paper --Vanilla Knockoffs, aggregated Knockoffs using e-values, aggregated Knockoffs using quantile-aggregation (AKO), KOPI and Knockoff inference via Closed Testing-- only Vanilla Knockoffs and KOPI yield discoveries, plotted above. All other methods are powerless. We use 50 Knockoffs draws, $\alpha = 0.1$ and $q = 0.2$. Each figure represents the region returned by a given method. Vanilla Knockoffs yield 32 regions, KOPI: 27 regions.}
\label{fig:fMRI7}

\end{figure}

The results corresponding to 7 contrasts of the HCP dataset are presented in Figs \ref{fig:fMRI1} -- Fig \ref{fig:fMRI7}:
\emph{foot} contrast of the HCP motor task in Fig \ref{fig:fMRI1},
\emph{hand} contrast of the HCP motor task in Fig \ref{fig:fMRI2},
\emph{relational versus match} contrast of the HCP relational task in Fig \ref{fig:fMRI3},
\emph{gain vs loss} contrast of the HCP gambling task in Fig \ref{fig:fMRI4},
\emph{2-back vs 0-back} contrast of the HCP working memory  task in Fig \ref{fig:fMRI5},
\emph{face vs shape} contrast of the HCP Emotional task in Fig \ref{fig:fMRI6},
\emph{interacting vs non-interacting} contrast of the HCP social task in Fig \ref{fig:fMRI7}.
These maps display the support of the conditional association test, with a sign that shows whether a region has an upward or downward impact on the decision function.

Overall, many Knockoff-based methods are powerless on all contrasts considered. Only KOPI, Vanilla Knockoffs and e-values aggregation consistently display non trivial solutions. This corresponds to the behavior observed in hard simulation settings in Fig. \ref{fig:sim1}, i.e. low SNR and high correlation for instance.

\end{document}